\documentclass[a4paper,11pt]{article}
\usepackage{jcappub}
\pdfoutput=1 

\usepackage{psfrag,color,epsfig}
\usepackage{graphicx,graphics}	
\usepackage{upgreek}
\usepackage{amssymb}	
\usepackage{multicol}	
\usepackage{bm}		
\usepackage{pdflscape}	
\usepackage{xspace}
\usepackage[T1]{fontenc} 
\usepackage{subfig}
\usepackage[multiple]{footmisc}
\usepackage{soul}
\newcommand{\HI}{$\mathrm {H{\scriptsize I}~}$}

\def\kk{{\mathbf\it{k}}}
\newcommand{\mc}[1]{{\color{black} #1}}
\newcommand{\mcb}[1]{{\color{black} #1}}
\newcommand{\mcbb}[1]{{\color{black} #1}}

\newcommand{\XHI}{{\rm x}_{\rm HI}}
\newcommand{\XHII}{{\rm x}_{\rm HII}}
\newcommand{\AVXHII}{\overline{{\rm x}}_{\rm HII}}

\definecolor{darkgreen}{rgb}{0, 0.70, 0}

\title{Inferring IGM parameters from the redshifted 21-cm Power Spectrum using Artificial Neural Networks}

\author[a,b,1]{Madhurima Choudhury, \note{Corresponding author.}}
\author[a,c]{Raghunath Ghara,}
\author[d,e,f]{Saleem Zaroubi,}
\author[f]{Benedetta Ciardi,}
\author[e] {Leon V. E. Koopmans,}
\author[g]{Garrelt Mellema,}
\author[a, h]{Abinash Kumar Shaw,}
\author[f]{Anshuman Acharya,}
\author[i]{I. T. Iliev,}
\author[j]{Qing-Bo Ma,}
\author[k]{Sambit K. Giri}

\affiliation[a]{Astrophysics Research Centre of the Open University of Israel, Ra'anana 4353701, Israel}
\affiliation[b]{Center for Fundamental Physics of the Universe, Department of Physics, Brown University, Providence 02914, RI, USA}
\affiliation[c]{Department of Physical Sciences, Indian Institute of Science Education and Research Kolkata, Mohanpur, WB 741 246, India}
\affiliation[d]{Department of Natural Sciences, The Open University of Israel, 1 University Road, Ra'anana 4353701, Israel}
\affiliation[e]{Kapteyn Astronomical Institute, University of Groningen, PO Box 800, 9700AV Groningen, The Netherlands}
\affiliation[f] {Max-Planck Institute for Astrophysics, Karl-Schwarzschild-Straße 1, 85748 Garching, Germany}
\affiliation[g]{The Oskar Klein Centre, Department of Astronomy, Stockholm University, AlbaNova, SE-10691 Stockholm, Sweden}
\affiliation[h]{Department of Computer Science, University of Nevada, Las Vegas, Nevada 89154, USA}
\affiliation[i]{Astronomy Centre, Department of Physics and Astronomy, Pevensey II Building, University of Sussex, Brighton BN1 9QH, UK}
\affiliation[j]{School of Physics and Electronic Science, Guizhou Normal University, Guiyang 550001, PR China}
\affiliation[k]{Nordita, KTH Royal Institute of Technology and Stockholm University, Hannes Alfvéns väg 12, SE-106 91 Stockholm, Sweden}
\emailAdd{madhurimachoudhury811@gmail.com}

\abstract{
The high redshift 21-cm signal promises to be a crucial probe of the state of the intergalactic medium (IGM). Understanding the connection between the observed 21-cm power spectrum and the physical quantities intricately associated with the IGM is crucial to fully understand the evolution of our Universe. In this study, we develop an emulator using artificial neural network (ANN) to predict the 21-cm power spectrum from a given set of IGM properties, namely, the bubble size distribution and the volume averaged ionization fraction. This emulator is implemented within a standard Bayesian framework to constrain the IGM parameters from a given 21-cm power spectrum.  
We compare the performance of the Bayesian method to an alternate method using ANN to predict the IGM parameters from a given input power spectrum, and find that both methods yield similar levels of accuracy, while the ANN is significantly faster. We also use this ANN method of parameter estimation to predict the IGM parameters from a test set contaminated with noise levels expected from the SKA-LOW instrument after 1000 hours of observation. Finally, we train a separate ANN to predict the source parameters from the IGM parameters directly, at a redshift of $z=9.1$, demonstrating the possibility of a non-analytic inference of the source parameters from the IGM parameters for the first time.  We achieve high accuracies, with R2-scores ranging between $0.898-0.978$ for the ANN emulator and between $0.966-0.986$ and $0.817-0.981$ for the predictions of IGM parameters from 21-cm power spectrum and source parameters from IGM parameters, respectively. The predictions of the IGM parameters from the Bayesian method incorporating the ANN emulator leads to tight constraints on the IGM parameters.
}

\begin{document}	
\maketitle
\flushbottom

\section{Introduction}
\label{intro}

The redshifted 21-cm line of neutral Hydrogen is a promising probe to study the evolution of our Universe. Over the first billion years, our Universe traversed three distinctive phases. The \textit{Dark Ages (DA)} marks a period when the Universe was dark and there were no luminous objects. The birth of the first stars and astrophysical objects ushers in the beginning of the \textit{Cosmic Dawn (CD)}. These first generations of sources began to heat and eventually ionize the neutral Hydrogen in the intergalactic medium (IGM). This period is called the \textit{Epoch of Reionization (EoR)}, and marks the last major phase transition in the history of our Universe. 
Observations of the redshifted 21-cm signal will enable us to study the origin and evolution of these first-generation sources
(e.g: \cite{Morales_2010, Pritchard_2012, Zaroubi_2013, Liu_2020}), and learn about the morphology and evolution of the ionized structures which were carved out by the first sources of light \cite[e.g.,][]{Furlanetto2016, 2018MNRAS.479.5596G, 2020MNRAS.496..739G, 2021JCAP...05..026K, 2024MNRAS.530..191G}. As the 21-cm signal is intimately connected to physical quantities describing the state of the IGM, these observations promise to open up a window to understand and interpret the evolution of the IGM with redshift. This, in turn, will enable us to better understand and constrain the properties of the sources and their evolution.  
The observations of the 21-cm signal power spectrum using large interferometric arrays currently hold the greatest potential to detect the redshifted 21-cm line \cite{Zaldarriaga_2004, Bharadwaj_2005, Morales_2005} and probe the large-scale distribution of neutral hydrogen across a range of redshifts. Detecting the 21-cm signal power spectrum is a major goal of several ongoing and future experiments, for example, the Giant Meterwave Radio Telescope (GMRT, \cite{Swarup_1991}), the Low Frequency Array (LOFAR, \cite{Harlem_2013_short}), Hydrogen Epoch of Reionization Array (HERA, \cite{DeBoer_2017_short}), the New Extension in Nançay Upgrading LOFAR (NenuFAR, \citep{Munshi2023}), the Amsterdam ASTRON Radio Transients Facility And Analysis Center (AARTFAAC, \citep{Gehlot_2022}),  the Murchison Wide-field Array (MWA, \cite{Tingay_2013_short}), 
have carried out observations to help constrain the 21-cm signal power spectrum from the CD and EoR. Other upcoming experiments such as the Square Kilometer Array (SKA, \cite{Mellema_2013, Koopmans_2015}) not only aim to measure the EoR 21-cm signal power spectrum with much-improved sensitivities, but even image the 21-cm signal directly \cite{ghara16}, promising to give a deeper insight into the physics of the evolution of the Universe.
Another set of experiments aim to measure the redshift evolution of the sky-averaged 21-cm signal or the global 21-cm signal. The Experiment to Detect the Global EoR Signature (EDGES, \cite{Bowman_2018}); Shaped Antenna measurement of the background RAdio Spectrum (SARAS, \cite{singh2021}); the Large-Aperture Experiment to Detect the Dark Ages (LEDA, \cite{Greenhill_2012}); the Radio Experiment for the Analysis of Cosmic Hydrogen (REACH, \cite{de_lera_Acedo_2022}) and the Cosmic Twilight Polarimeter, (CTP, \cite{Nahn_2018}) are examples of such experiments. 
 
The 21-cm signal is faint compared to other astrophysical signals, hence detecting it is very challenging. One of the major observational challenges is the foregrounds, comprising of galactic and extra-galactic components which are about 3 to 4 orders of magnitude brighter than the 21-cm signal. The earth's ionosphere, radio-frequency interference(RFI) and the instrument response also contaminates all ground-based observations of the 21-cm signal, making its detection dependent on the accuracy of RFI and foreground removal, instrument calibration methods and sensitivities \cite[]{Datta_2010a, barry_2016, Liu_2020, Hothi_2021, Kern_2019ApJ, 2020Mevius, Gan_2023, 2025arXiv250321728M}. All experiments that aim to detect the faint \HI signal aim to improve and develop better signal extraction and calibration algorithms. In recent works, we have been obtaining more stringent upper limits on 21-cm signal power spectrum measurements from the EoR, with increasingly more sensitive instruments and better calibration methods. \mc{For example, the recent upper limits from LOFAR, HERA  and MWA are presented in \cite{Mertens_2020, 2025arXiv250305576M},\cite{Zara_2022ApJ_short, Keller_2023_short}, \cite{Trott2020MNRAS.493.4711T} respectively. }

Once we have the measurements of the 21-cm signal, we aim to understand what kind of theoretical models could explain the measurements, and constrain the associated astrophysical and cosmological parameters. For the analysis of the 21-cm power spectrum, a large suite of simulations are required which represents different reionization scenarios. These are used to constrain the properties of the source parameters, for example: the escape fraction of UV photons, star formation efficiency, etc. Also, using upper limits obtained from the experiments, one can rule out certain reionization scenarios. For example, \cite{Ghara_2020, Greig_lofar, 2020MNRAS.498.4178M, 2021MNRAS.503.4551G, ghara2025} used a suite of simulations and a Bayesian method to constrain the source and IGM parameters, and were able to rule out a range of cold IGM models using the LOFAR upper limits obtained in \cite{Mertens_2020}. To extract useful information from the 21-cm observations, one can use parameter estimation techniques such as the Markov Chain Monte Carlo (MCMC) \cite[for example,][]{Greig_2015, 2022JCAP...03..055G, Ghara_2024} or Artificial Neural Networks (ANN) \cite{Shimabukuro_2017, Choudhury_2020, Choudhury_2022, 2024arXiv241216853G}.  

In recent years, machine learning (ML) techniques have been increasingly used in various aspects of cosmology, astrophysics, statistics, inference, and imaging. Particularly in the field of 21-cm cosmology and inference, there have been several applications of ML algorithms \cite{2024MNRAS.527.7835A}. \mcb{One of the most common applications are in the development of fast emulators for 21-cm global signal \cite{Cohen_2019, Sikder_2024,Bye_2022, Bevins2021}, power spectrum \cite{Schmit_2017, Kern_2017, Breitman_2024} and bispectrum studies \cite{tiwari2021}. In \cite{Hassan_2019}, the authors use Convolutional Neural Networks (CNN) to differentiate between 21-cm images produced by different models of reionization, while in \cite{Bianco2021}, the authors use deep learning to identify ionized regions in 21-cm maps.} Deep learning methods have been used to emulate the entire time evolving 21-cm brightness temperature maps from the EoR as well \cite[as in ][]{Chardin_2019}. More recent implementations of ML include convolutional denoising autoencoder (CDAE) to recover the 21-cm signal from the EoR by training on SKA images simulated with realistic beam effects \cite{Li_2019}.\cite{Shimabukuro_2022} used ANN to estimate bubble size distribution from various 21cm power spectra. CNN have been used on 3D-tomographic 21-cm images for parameter estimation and posterior inference \cite{Zhao_2022}. \mcb{Machine learning methods have also been implemented in \cite{Gagnon-Hartman2021, Kenedy_2024} to recover modes lost to 21-cm foregrounds.} \cite{Choudhury_2022} used ANN for extracting the parameters of the 21-cm power spectrum from foreground dominated synthetic observations. 

While the 21-cm signal offers valuable insights into the IGM properties, it doesn't directly reveal the properties of the sources responsible for its ionization, the intricate interplay of feedback processes, and other complex astrophysical phenomena. Inferring the source properties from the 21-cm power spectrum directly, heavily relies on assumptions about intricate physical and astrophysical processes. 
Even when we have a solid understanding of the source-IGM relationship, the resulting differences in the observable 21-cm brightness temperature fluctuations $(\delta T_b)$ can be subtle and difficult to discern. This inherent degeneracy, exacerbated by the limitations of current low-frequency telescopes, often allows a range of source models to explain the same observed $\delta T_b$. As the 21-cm power spectrum primarily reflects the characteristics of the IGM, inference methods based solely on the IGM parameters would give us a clearer understanding of the 21-cm power spectrum. 

Traditional 21-cm inference frameworks have mostly prioritized constraining source properties. This paper, however, champions the direct measurement of IGM properties, unveiling a framework solely dedicated to interpreting the 21-cm power spectrum's connection to the IGM.  The initial framework is based on a single redshift at $z = 9.1$ (which is one of the LOFAR redshifts), and we plan to develop a multi-redshift framework in the future. \mc{ In \citep{Mirocha_2022}, the authors have presented a phenomenological galaxy-free model for 21-cm fluctuations during reionization that works directly in terms of the mean properties of the IGM and the size distribution of ionized regions, considering uniform spin temperature field, spherical non-overlapping ionized bubbles and a binary ionization field. Recently, \citep{Ghara_2024} described an ansatz consisting of a set of redshift and scale- independent parameters to model the redshift evolution of the ratio of the 21-cm power spectrum from the EoR and the corresponding density power spectrum. }

The paper is organized as follows. In \textsection~\ref{section:theory}, we describe the 21-cm power spectrum and the details of the simulation and the models for the 21-cm signals used.  In \textsection~\ref{section: ann-emu}, we briefly introduce the concept of Artificial Neural Networks(ANN), followed by a detailed description of the design of the emulator. In \textsection~\ref{section: results}, we discuss the results obtained, the implications of our findings in  \textsection~\ref{discussions} and finally conclude in section \textsection~\ref{conclusions}.  Throughout the paper, we adopt the background cosmological parameters as  $\Omega_m=0.27$, $\Omega_\Lambda=0.73$, $\Omega_b=0.044$, $h=0.7$  \cite[WMAP;][]{2013ApJS..208...19H}. These cosmological parameter values were used in the $N$-body simulations employed here.

\section{Theory}
\label{section:theory}
In this section, we briefly introduce the 21-cm signal and the power spectrum, followed by a detailed description of the modelling of the 21-cm signal with the 1D radiative code {\sc grizzly}.

\subsection{The \HI 21-cm signal and power spectrum}
\label{subsection: 21cm signal theory}
The hyper-fine splitting of the lowest energy level of hydrogen, gives rise to the rest-frame frequency $\nu_{21} = 1.42$ GHz radio signal, corresponding to a wavelength of about 21 cm. The 21-cm signal is produced by the neutral hydrogen atoms in the IGM, collectively observed against the background CMB radiation (\cite{Madau_1997, Pritchard_2012, Zaroubi_2013} give a detailed review). This resulting contrast is the differential brightness temperature, which is the primary observable in 21-cm experiments, given by: 

\begin{equation}
\begin{split}
      \delta T_{b}(\textbf{x},z) \approx ~ & 27~{\XHI(\textbf{x},z)\left[1+\delta_{b}(\textbf{x},z)\right]}\left(\frac{\Omega_{b}h^{2}}{0.023}\right) \\ &
      \left(\frac{0.15}{\Omega_{m}h^{2}} \frac{1+z}{10}\right)^{1/2} 
      \left[1-\frac{T_{\gamma}(z)}{T_{s}(\textbf{x},z)}\right]\Big[\frac{\partial_{r} v_{r}}{(1+z)H(z)}\Big]^{-1},
\label{eq:global}
\end{split}
\end{equation}
where, $\XHI$ is the neutral fraction of hydrogen, $\mathrm \delta_{b}$ is the fractional over-density of baryons, $\mathrm \Omega_{b}$ and $\mathrm \Omega_{m}$ are the baryon and total matter density, respectively, in units of the critical density. $H(z)$ is the Hubble parameter and $T_{\gamma}(z)$ is the CMB temperature at redshift $z$, $T_{s}$ is the spin temperature of neutral hydrogen, and $\partial_{r} v_{r}$ is the velocity gradient along the line of sight. 
The 21-cm power spectrum, $P_{21}(k)$, from the EoR can be measured using the fluctuations of the brightness temperature \mc{field, $\delta T_b(\mathbf{x},z)$. It is defined as:  }
\mc{
\begin{equation}
        \rm{<\delta T_b(\bar \kk)~\delta T_b^*(\bar \kk')>}=\rm{ (2\pi)^3\delta(\kk-\kk')P_{21}(\bar \kk)}
\end{equation}
where ⟨...⟩ denotes an ensemble average, $\delta T_b(k)$ is the
Fourier transform of $\delta T_b(x)$ and $\delta_D$ is the Dirac delta function.}
The dimensionless power spectrum of the brightness temperature is given by ${\Delta^2=\kk^3P_{21}(k)/2\pi^2}$. Throughout this paper, we work with the dimensionless power spectrum, ${\Delta^2}$, in units of $\mathrm{mK^2}$.
\begin{figure*}
    \centering
    \includegraphics[width=\textwidth]{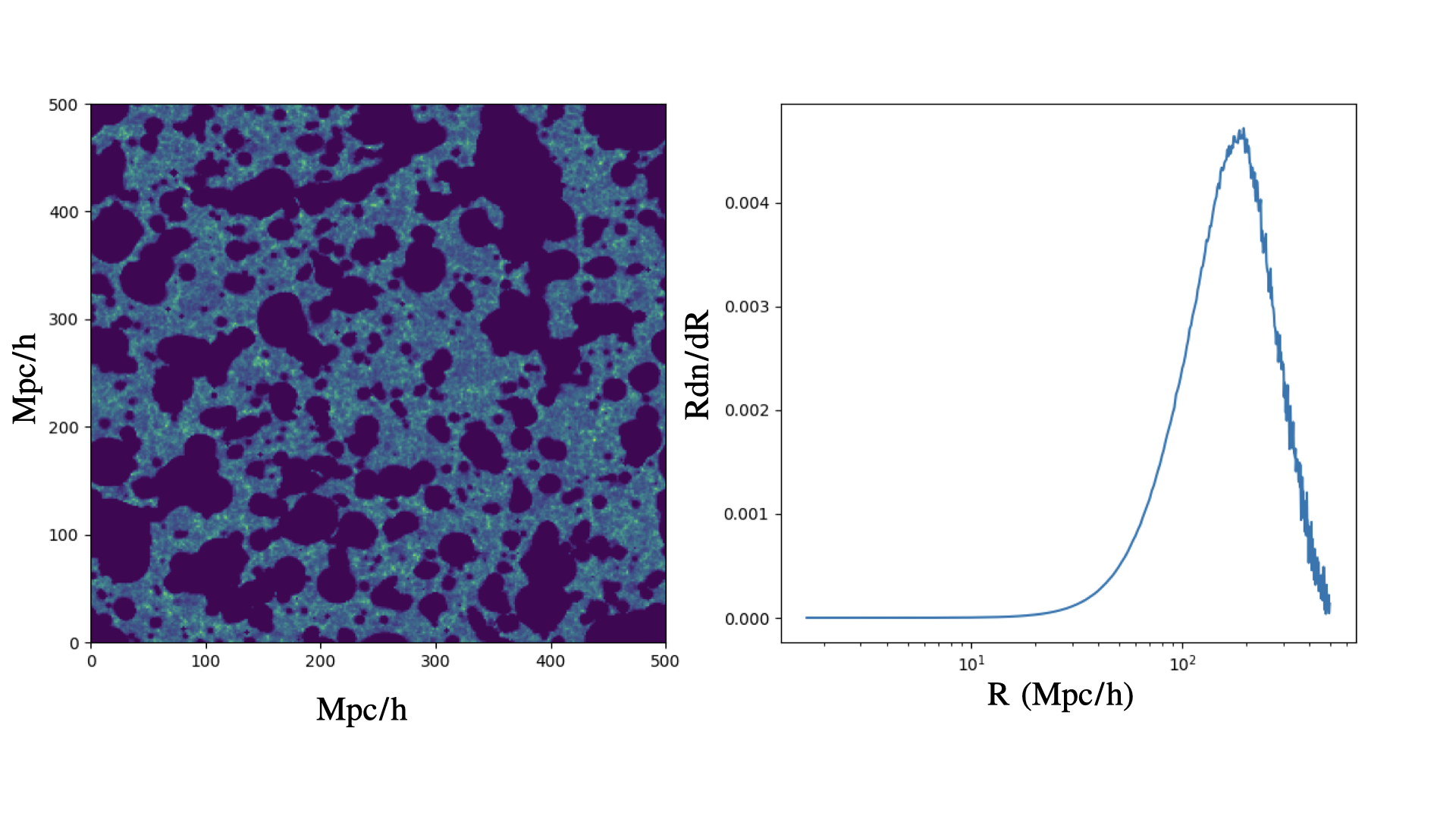}
    \caption{In the left panel, we show a slice from a {\sc grizzly} simulation at $z= 9.1$. The solid dark regions are the completely ionized regions. This slice has a volume averaged ionized fraction of  $0.40$.  The plot on the right is a size distribution of ionized bubbles obtained from the ionized map using the mean-free-path method.}
    \label{fig:slice-bsd}
\end{figure*}

\subsection{Modeling the 21-cm signal with {\sc grizzly}}
\label{subsection: grizzly source models}
In this paper, we use the one-dimensional radiative transfer code, {\sc grizzly}  \cite[][]{Ghara_2015, ghara18} to model the 21-cm signal. The algorithm uses uniformly gridded dark-matter density cubes and their corresponding dark-matter halo-lists, along with the velocity-field cubes as inputs. Given a source model, {\sc grizzly} then generates the ionization maps and the brightness temperature maps during the EoR. In our models, we assume $T_s \gg T_{\gamma}$, which is expected to be the scenario in the presence of efficient X-ray heating \citep[see e.g,][]{Pritchard07, Mesinger_2011, 2019MNRAS.487.2785I, Ross2019}. This implies that the large-scale fluctuations of the signal are mostly driven by density and $\XHI$ fluctuations.  For this paper, the simulations considered use the gridded dark-matter density fields and
the corresponding halo lists within comoving cubes of side 500 $\mathit{h}^{-1}$ Mpc, produced from the PRACE4LOFAR\footnote{Partnership for Advanced Computing in Europe: \url{http://www.prace-ri.eu/}} N-body simulations (for the details of the simulation, see e.g, \cite[]{2019JCAP...02..058G, 2021MNRAS.502.3800K}).  We vary the following parameters, which control the source models to generate a suite of simulations:
 \begin{enumerate}
     \item Ionization efficiency $\zeta$: The rate of ionizing photons per unit stellar mass, escaping from a halo is given by, $\Dot{N_i} = \mathrm{\zeta\times 2.85 \times 10^{45}~ s^{-1}~M_{\odot}^{-1}}$. Here, $\zeta$ is the ionization efficiency. The rate of ionizing photons is assumed to be proportional to the halo mass and the ionizing efficiency, $\zeta$ combines the degeneracies from other astrophysical parameters (such as the star formation rate, the emission rate of ionizing photons from the sources, as well as their escape fraction) in the IGM.
     \item Minimum mass of the UV emitting halos ${M_\mathrm{min}}$: The number of ionizing photons escaping from a halo is assumed to be linearly dependent on its mass. However, below a certain minimum mass, radiative and mechanical feedback can severely reduce the star formation efficiency \cite{Hasegawa_2012, Dawoodbhoy_2018}. We thus introduce ${M_\mathrm{min}}$, which is the minimum mass of the halos, from which ionizing photons escape into the IGM.
\end{enumerate}

For a set of source parameters $\boldsymbol{\theta}=[{\zeta, M_\mathrm{min}}]$ and the previously described inputs from the N-body simulation, {\sc grizzly} first generates ionization maps at the given redshift. These $\XHII$ maps are later used to produce coeval $\delta T_b$ maps using Eq. \ref{eq:global}. Note that we use the cell moving technique \cite[][]{ghara15b, 2021MNRAS.506.3717R} to include the impact of the redshift space distortion.  

\subsection{Characterizing the IGM}
\label{subsection: igm description}
Most EoR simulations are dependent on the source parameters, a combination of which can be used to generate several reionization scenarios. However, as the 21-cm signal is a more direct probe of the state of the IGM, we choose to characterize the IGM with the following parameters. 
\begin{enumerate}
    \item The volume-averaged ionization fraction, $\AVXHII$: As the Universe transitioned from a predominantly neutral state to an ionized state, the ionized regions gradually began to grow in size. This led to the formation of a network of ionized regions, with ionized bubbles surrounding dense regions hosting ionizing sources. The volume-averaged ionization fraction, $\AVXHII$  provide a measure of the ionization state of the Universe, at a certain redshift. 
    \item The bubble size distribution (BSD): The size distribution of regions containing ionized Hydrogen during the EoR is an important quantity which can encapsulate crucial insights about early stars and galaxy formation. At a given redshift, the tomography of the IGM can be best described by such a size distribution. The BSD describes the range of sizes that these ionized regions can have, and their relative abundances. There are a few methods of determining the size distribution of the bubbles, for example, the Friends-of-Friends (FoF) method and the Mean-Free-Path (MFP) method. The FoF method is based on the percolation theory and works well when there are only bubbles of equal sizes, focussing on the connectedness of the ionized regions. The FoF method is computationally very expensive and fails when there are bubbles of different sizes, particularly in the later stages of reionization when there are several overlapping ionized regions.  In this work, we use the MFP method to determine the size distribution of the bubbles. To obtain a bubble size distribution from a simulation, a pixel within an ionized region is randomly chosen, and the distance from this point to the nearest neutral pixel in a random direction is measured and recorded. This procedure is repeated several times $(\sim 10^7$ times), and these distances are recorded, following which a probability distribution function (PDF) of these distances is obtained. The bubble size distribution can be derived by taking the volume-weighted average (\cite{Mesinger_2007}) of the probability distribution function. The PDF is given by $\mathrm{R\frac{dn}{dR}}$,
    where, n denotes the number of recorded lengths or rays or mean-free-paths, in the range from $\mathrm {R}$ to $\mathrm{R + dR}$. In Fig.\ref{fig:slice-bsd}, we show a slice from a {\sc grizzly} simulation with a volume averaged ionization fraction of $\sim 40\%$. The right-hand panel of the figure shows the corresponding bubble size distribution. It should be noted that as a direct implication of the fact that Gaussianity is driven by large number statistics, the mean of the distribution is driven by the bubbles at larger scales. 
\end{enumerate}
These IGM parameters are derived quantities, and are very informative in characterizing the state of the IGM at a specific redshift. Nonetheless, while considering the dynamic evolution of the IGM over a range of redshifts, additional parameters would be important to comprehensively capture its characteristics.

\section{Artificial Neural Networks}
In this section, we describe the methodology to develop our ANN-based framework that connects the IGM-related quantities to the 21-cm signal power spectrum. We describe the basic architecture of the ANN and elaborate on the details of how the training and test datasets were created to develop our framework. 

\label{section: ann-emu}
\subsection{Brief overview of ANN}
Artificial Neural Networks (ANN) are one of the several available machine learning techniques which can be used for developing supervised learning-based frameworks. A neuron is the most functional unit of an ANN.
A basic neural network has three kinds of layers: an input layer, one or more hidden layers, and an output layer. In a fully connected network, each neuron in a layer is connected with every neuron in the next layer, and the connection is associated by a weight and a bias. ANNs are capable of mapping associations with a given input data and associated target parameters of interest, without the knowledge of their explicit causal relationships. During the training process, the network gradually optimises a chosen cost function by repeatedly back-propagating the errors and adjusting the weights and biases. A chunk of the data is put aside as the validation dataset and the performance of the network is validated using this set. For a detailed description of a similar training process, please see \cite{Choudhury_2020}. Once training and validation are completed, a test set is used to check the robustness of the network and predict the output parameters. For a detailed mathematical description of how the weights and biases are adjusted for every iteration, see \cite{Bishop_2006},  also summarized in \cite{Choudhury_2020}.\\

\begin{figure}
\centering    \includegraphics[width=0.85\linewidth]{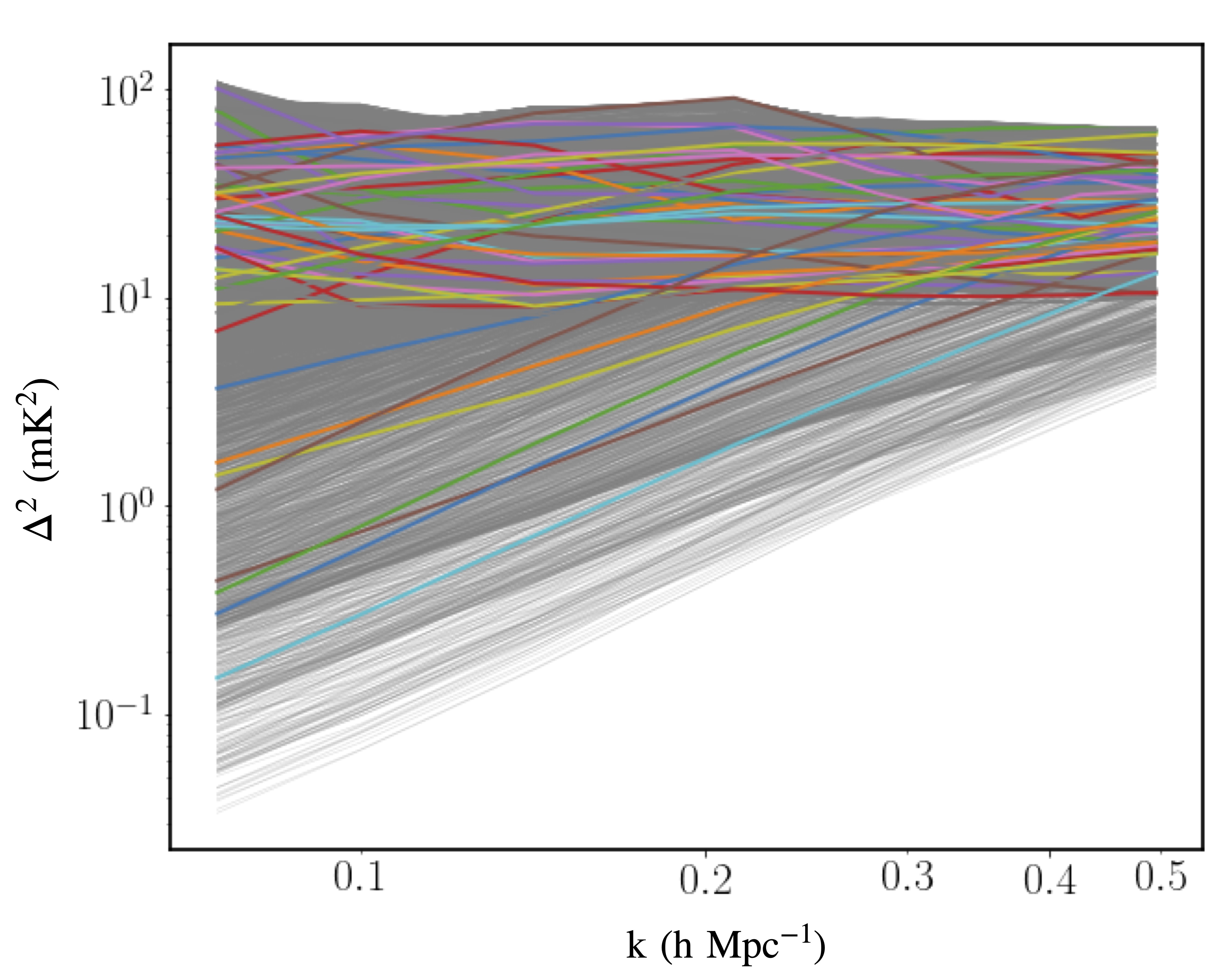}
    \caption{This is the power spectrum set consisting of 17000 models used for training the ANN. In this plot, a fraction of the training set is plotted in color, while the rest have been plotted in grayscale, to provide an idea of the different shapes of the power spectra included in the training set.}
    \label{fig:trainset}
\end{figure}

Training of ANN in a supervised manner is very target-specific, and can have very different network architectures for different applications. While ML methods are extremely fast and computationally very efficient, the ANN model performances need to be carefully analysed. There can be issues of ``overfitting'', which is typically the case when the trained model fails to generalize what it has learned during the training process. Such a scenario can be overcome by suitable regularization techniques. There can also be an ``underfitting'' scenario, which can be identified if the training error does not reduce even after several iterations. This can also be solved by making the network more complex (e.g: by adding more layers) and also by using a well-represented training dataset. \\

Once the ANN is trained, validated and tested, one requires to define a metric to quantify the performance and robustness of the network. There are various available metrics which can be computed to provide a quantification of the performance of the ANN, of which the most commonly used metrics for supervised learning methods are RMSE (root mean squared error) and R2 score (also called the coefficient of determination), which provide a measure of error in the predictions of the ANN. In this paper, we compute R2-scores for each of the output parameters as the performance metric. The R2-score is defined as:

\begin{equation}
\mathrm{R^2=1-\frac{\Sigma(y_{pred}-y_{orig})^2}{\Sigma(y_{orig}-\overline{y}_{orig})^2}}
\end{equation}
where, $\mathrm{{y}_{orig}}$ is the original parameter, $\mathrm{y_{pred}}$ is the parameter predicted by the ANN, $\mathrm{\overline{y}_{orig}}$ is the average of the original parameter, and the summation is over the entire test set. $\mathrm R^2$ can vary between 0 and 1, while $\mathrm R^2 =1$ implies a perfect inference of the parameters.

\subsection{Preparing the training and test sets for the 21-cm PS emulator}
\label{subsection: preparing training sets}
We have used publicly available python-based packages \textsc{scikit learn} (\cite{Pedregosa_2011}) and \textsc{keras} in this work to design the networks.
In this paper, we use two separate sets of simulations. 

\begin{itemize}
    \item Training set I: In our effort to span most of the parameter-space of the IGM and various shapes of the 21-cm power spectrum, we modify the source model described in \textsection\ref{subsection: grizzly source models}. We describe a source model in which the rate of ionizing photons escaping from a halo is given by, $\Dot{N_i} = \mathrm{\mathcal{A} \cdot \mathcal{M}_{halo}^{\alpha}}$. Here, $\mathrm {\mathcal{A}}$ is a proportionality constant, which includes other source parameters and $\alpha$ determines the non-linear dependency of the number of ionizing photons on the mass of the halo. This aims to encompass various scenarios of reionization, incorporating different morphologies that could result in different size distributions. \mc{We prepare the training set by varying $\alpha\in (-2,2)$. Following this, we grid the ionization fraction, $\XHII\in(0.20,0.90)$ uniformly and sample from the suite of simulations} (Fig.\ref{fig:trainset}).  To avoid any bias in our training set due to over-representation of any particular $\XHII$, we ensure that $\XHII$ is sampled uniformly in the range $\in(0.20,0.90)$, resulting in $\sim 17000$ models which we use to construct the training set. Thus, our training dataset comprises of the ionized fraction and the normalized bubble size distribution for each of the model, and the corresponding 21-cm power spectra. We generate a test set consisting of 1466 samples following the same method, and call it the {\sc grizzly} test set. This training set is used to train the emulator (ANN-Emu) (\textsection~\ref{results:grizzly}) and the ANN for the IGM parameter estimation (ANN-IGMParam) (\textsection~\ref{results:param est}).
    
    \item Training set II: The second set of training samples is generated using the source model as described in section \textsection~\ref{subsection: grizzly source models}. We use simulations with a box size of $\rm 500~\mathit{h}^{-1} \mathrm{Mpc}$ and compute the corresponding 21-cm power spectra, bubble size distributions and the volume averaged ionization fraction for each model. \mc{For this training set, we have fixed $\alpha = 1$ and varied the source parameters so as to obtain different reionization scenarios with different $\AVXHII$. This is specifically done by choosing a particular $M_{min}$ within the range, and then tuning the parameter $\zeta$ such that we obtain roughly a particular $\AVXHII$. In this way, we fill up the parameter space corresponding to $\AVXHII$.  We then grid and select samples from the $\AVXHII$ parameter space uniformly to create a well represented Training set II. It is to be noted that we do not enforce a uniformly gridded source parameter space $[\rm \zeta, M_{min}]$ for training this model}. We simply record the source parameters for this set, which we use later in \textsection\ref{results: source from IGM}.
    This set is smaller in size, comprising of $\sim 7600$ samples. This training set is used to train an ANN to determine source properties $[\rm \zeta, M_{min}]$ directly from the IGM properties $[\mathrm{BSD, \AVXHII}]$ (ANN-Source), which we discuss in detail in \textsection~\ref{results: source from IGM}.
\end{itemize}

We describe the architecture and the results from the ANN emulator and parameter estimation in the following section. We have summarized the different ANNs, their target and their corresponding training set in Tab.\ref{Tab:1}. \mcb{For each of the neural networks, we closely monitor the learning curves, particularly the loss and model accuracy curves for the training and validation sets. We also look at the performance metrics closely to ensure there is no overfitting. }

\begin{table}
\centering
\setlength{\tabcolsep}{4pt} 
\renewcommand{\arraystretch}{0.9} 
\begin{tabular}{|p{3.5cm}|p{5cm}|p{3cm}|}
     \hline
     \textbf{ANN name} & \textbf{Description} & \textbf{Training set} \\
     \hline
     ANN-Emu & Emulator to predict 21-cm PS from BSD and $\AVXHII$ & Training set I\\
     ANN-IGMParam & IGM Parameter estimation, $[A, \mu, \sigma, \AVXHII]$ from 21-cm PS  & Training set I\\
     ANN-Source & Source parameter estimation, $[\rm \zeta, M_{min}]$ from $[A, \mu, \sigma, \AVXHII]$  & Training set II\\
     \hline
\end{tabular}

\caption{ List of the different ANNs and their corresponding training sets }
\label{Tab:1}
\end{table}
\section{Results}
\label{section: results}
We present the results from the ANN emulator(ANN-Emu) for {\sc grizzly} test set. This emulator is implemented into a Bayesian framework to constrain the IGM parameters. Subsequently, we present the results of IGM parameter estimation using a trained ANN, and compare the results with the Bayesian framework. We further use noisy test datasets corresponding to 1000 hours of observation of the SKA-low configuration to predict the IGM parameters. Finally, we determine the source parameters from the IGM parameters using an ANN.

\subsection{Predictions of the 21-cm power spectra from {\sc grizzly} test sets }
\label{results:grizzly}
\begin{figure*}
    \includegraphics[width=\linewidth]{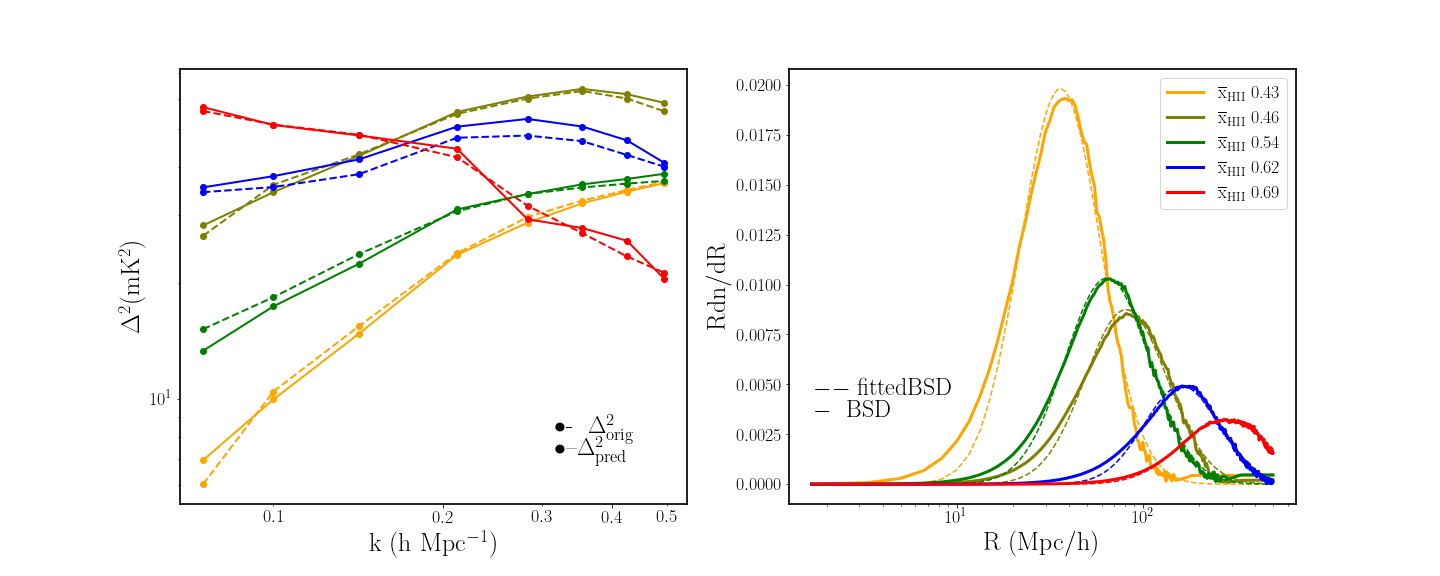}
    \caption{The panel on the left side shows the predictions of the ANN, from a random subset of the test set which comprises of {\sc grizzly} models at redshift 9.1. The plot shows the true and the predicted power spectra from the ANN emulator, represented by the solid and the dotted curves respectively. The solid curves on the right show the corresponding BSD and their ionization fractions. The dashed curves on the right panel are the log-norm fitted BSDs. }
    \label{fig:GrizzlyANN}
\end{figure*}

For the ANN emulator, we use a Multilayer-perceptron based ANN with 4 hidden layers, and activation functions `elu' for each of hidden layers. The number of input neurons corresponds to the dimension of the input dataset, which is 301 in our case (300 for the BSDs and 1 for the $\AVXHII$, per realization). The output layer has a linear activation function, and provides the 21-cm power spectrum for a range of $\mathit{k}$-values as the output. \\

\begin{figure*}
    \centering
    \includegraphics[width=\textwidth]{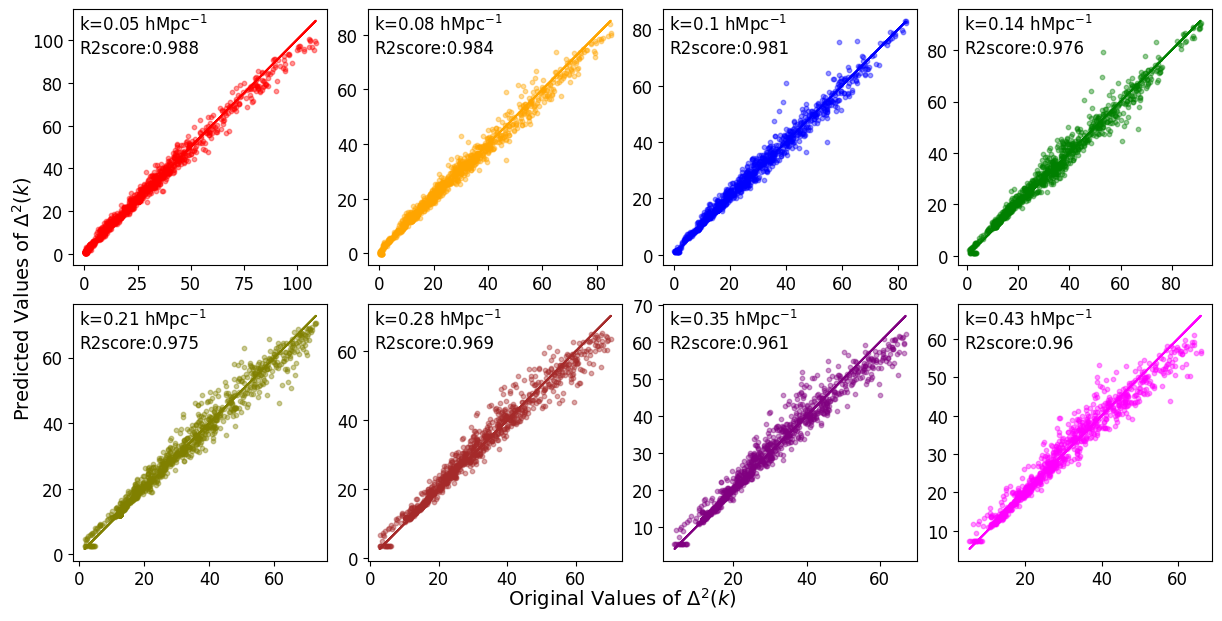}
    \caption{ The true versus predicted values from the {\sc grizzly} test set from ANN-Emu, for each k-mode along with their R2 scores are shown in this figure.  }
    \label{fig:predictions}
\end{figure*}

\begin{figure*}
    \centering
    \includegraphics[width=\textwidth]{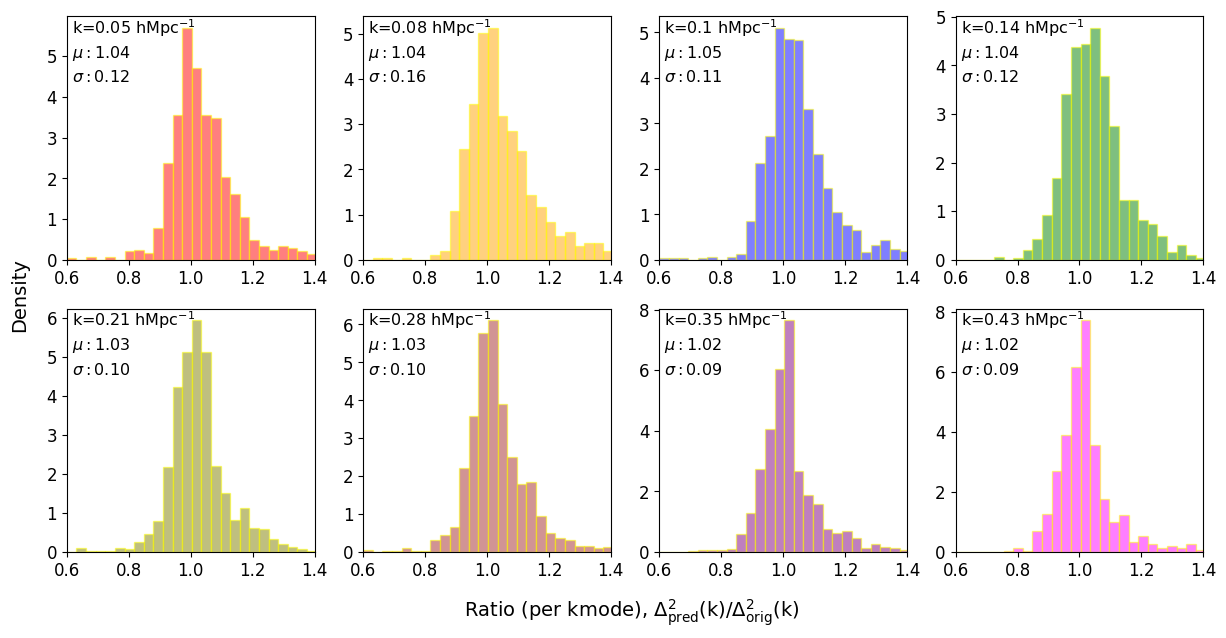}
    \caption{ The ratio of the true versus the predicted values from the {\sc grizzly} test set for each k-mode are shown in this figure, along with the mean $(\mu)$ and standard deviation $(\sigma)$ of these distributions. }
    \label{fig:predictions_histogram}
\end{figure*}

To assess the performance of the ANN emulator, we first consider the {\sc grizzly} test set which contains  1466 samples. Given a BSD, along with its ionized fraction at redshift 9.1, the emulator is trained to predict the 21-cm power spectrum for the $\mathit{k}$-modes [0.05, 0.08, 0.1, 0.14, 0.21, 0.28, 0.35, 0.43] $\mathrm {\mathit{h}Mpc^{-1}}$.  In Fig.\ref{fig:GrizzlyANN}, we show the predictions from the ANN emulator for a fraction of the test set consisting of models using the {\sc grizzly} simulation (or the {\sc grizzly} test set). In the right panel of the same figure, we plot the corresponding BSD along with the ionized fractions (mentioned in the legend). 
In order to better demonstrate the accuracy of predictions from our ANN model, we present scatter plots of the $\mathrm {\Delta^2_{pred}}$ versus $\mathrm {\Delta^2_{true}}$ per k-bin with the corresponding R2-scores in Fig.\ref{fig:predictions}, where $\Delta^2$ is the dimensionless power spectrum and the subscripts "true" and "predicted" denotes the true 21-cm power spectrum and the predicted 21-cm power spectrum respectively. If we plot the distribution of the ratio of true versus predicted power spectra per k-bin (see Fig.~\ref{fig:predictions_histogram}, we observe that, for more than $95\%$ of test set samples,  the distribution is centered around 1. 

Although the BSD is a reasonably good statistical representation of the distribution of the sizes of the ionized bubbles, there is some information lost while transitioning from the ionized maps to the size distributions, particularly in scenarios where there are complex shaped, overlapping structures in advanced stages of reionization. From the scatter plots in Fig.\ref{fig:predictions}, we see that the R2-scores from the predictions for different $\mathit{k}$-values are between $0.898-0.978$. An R2-score closer to 1 implies a very good prediction. We also observe that at higher k-values $\mathrm{(\mathit{k}>0.08 ~\mathit{h}Mpc^{-1})}$, the R2-score is not as good as compared to the two lowest $\mathit{k}$-modes we consider in this work. 
\mc{The training set consists of a large variety of reionization morphologies, including pre-overlap and post-overlap reionization stages. The BSD lacks information of these spatial locations, shapes, and clustering of the ionized regions and hence it is not straightforward to infer the reasons for the comparatively poor constraints at smaller scales. These comparatively less accurate predictions at the larger k-values(smaller scales), could be due to the fact that there are comparatively fewer models in the entire training set which specifically capture the correlations between the smaller scales and the 21-cm power spectrum.} 
\mc{We have  performed hyperparameters tuning using keras-tuner to obtain the best combination of hyperparameters for each of the ANN models. We would like to stress that the performance of the ANN model is sensitive to the fact that the current choice of the set of the IGM parameters does not, in principle, completely describe the IGM. That is reflected in the network performance.
}
We will be incorporating this emulator in the following section to obtain Bayesian constraints on the IGM parameters. 

\begin{figure*}
    \centering
    \includegraphics[width=1.1\textwidth]{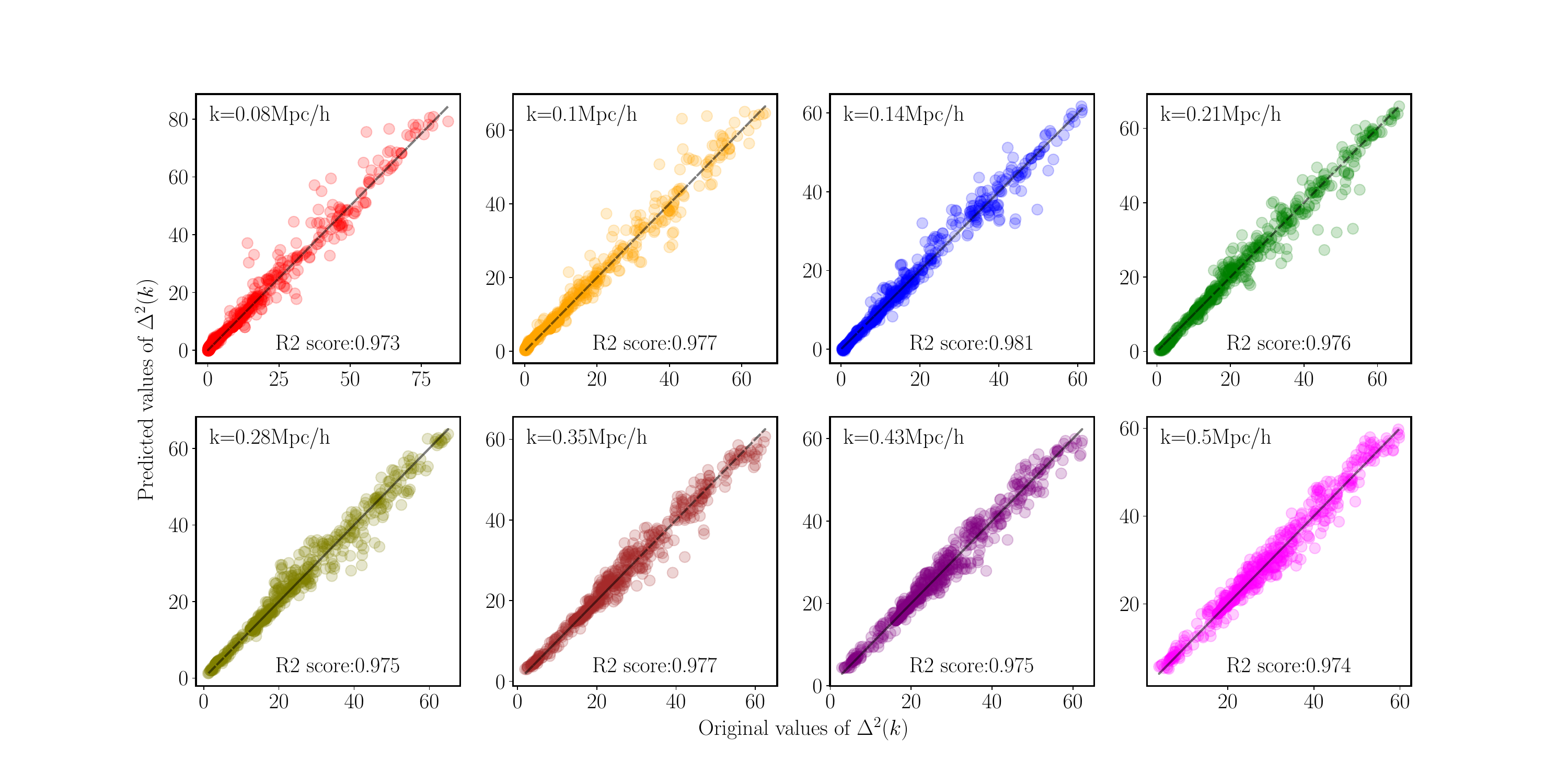}
    \caption{\mcb{Predictions from the NN trained on the very early stages of reionization, when there wasn't any overlapping bubbles. These results show how the larger k-modes(corresponding to smaller scales) are better constrained than the smaller k-modes at these early stages of reionization.}}
    \label{fig:pre-overlap}
\end{figure*}
\mcb{It is important to note that the power spectrum and state of the IGM at very early stages of reionization is quite different. Training set I consists of various different morphologies ranging between volume averaged ionization fraction, $\AVXHII \in (0.20,0.90)$. We train a separate NN to understand the association between the morphology of the IGM and the 21-cm power spectrum, at very early stages of reionization $(\AVXHII<0.20)$, when there are small bubbles and no overlapping at all. We plot the true vs predicted values of the power spectrum from the results obtained by training on the pre-overlap dataset in Fig.\ref{fig:pre-overlap}  Comparing these plots with the results from training on the entire dataset, we observe that, at the lower k-values (large scales) the NN is less efficient in constraining the 21-cm power spectrum. This is expected, as the BSD does not have enough information at these scales in the early stages of reionization. If we look at the 21-cm power spectrum predictions at higher k-values(smaller scales), we see they are much better constrained as compared to the lower k-modes. This demonstrates the importance of extent of the largest ionized regions and hence the dominant scales in the reionization morphology in the power spectrum measurements. }


\subsection{Bayesian constraints on IGM parameters using the trained ANN model}
\label{results:bayesian}
For a given model, placing robust and quantitative constraints on the model parameters (given a series of observations and their associated uncertainties) is the basic motivation of parameter inference. A posterior probability distribution $\mathcal{P}(\theta |x)$ of the model parameters, $\theta=\{\theta_1,\theta_2,...,\theta_n \}$, given a data vector, $x=\{ x_1,x_2,...,x_n\}$, is usually computed following Bayes' theorem:
\begin{equation}
    \mathcal{P}(\theta|x)=\frac{\mathcal{L}(x|\theta)\pi(\theta)}{P(x)},
\end{equation}
where, $\mathcal{L}$ is the likelihood function of the data, $\pi(\theta)$ is the prior probability distribution of the model parameters, and $P(x)$ is the probability distribution of the given data. The maximum likelihood estimate is found by maximizing $\mathcal{L}(x|\theta)$, and the maximum a posteriori estimate is computed by maximizing $\mathcal{P}(\theta|x)$. This entire process works well as long as the parameter space is not very vast, and the parameters themselves are also well-behaved.

In this subsection, we present the Bayesian constraints on the IGM parameters. We do so, by using the ANN emulator (described in detail in \textsection~\ref{results:grizzly}), which has been trained to predict the 21-cm power spectrum from a given bubble size distribution and the ionized fraction. \mc{We assume a Gaussian likelihood function, :
\begin{equation}
    \log \mathcal{L}=-0.5\cdot  \left(\sum\frac{(y_{model}-y)^2}{y_{err}^2}\right).
\end{equation}
This specific likelihood assumes that the observed data y
has Gaussian-distributed errors around the model predictions $y_{model}$, with 
$y_{err}$ representing the standard deviation of these errors for each data point. In this case, the negative log-likelihood is proportional to the sum of squared residuals, weighted by the inverse of the variance (i.e., $y_{err}^2$ ), which accounts for how well the model predictions fit the data under Gaussian-distributed uncertainties. \mcb{Ideally, the total error budget should incorporate contributions from measurement uncertainty, sample variance, model imperfections, and emulator errors to provide a more comprehensive characterization of uncertainties. \mcbb{We have assumed a very simple case where the error is not dependent on {\it{k}} or $z$ in our likelihood function, $y_{err}= 0.2 \rm mK^2$ for the current Bayesian analysis. }We have not incorporated any instrument or emulator error in our assumption of $y_{err}$ in the likelihood function, and plan look into more realistic error propagations in a future work.
} }

If we directly use the ANN-emulator as the model for Bayesian inference, we would require 301 parameters in total, corresponding to the number of inputs that goes into the ANN (the first 300 corresponding to the BSD and one corresponding to the ionized fraction). That would result in a very large number of parameters to constrain. To avoid this, we opt for a simpler parametrization of the bubble size distribution as a log-normal distribution (Eq.\ref{eq:BSDparam}), where, A, $\mu, \sigma$ are, respectively, the amplitude, mean and standard deviation of the corresponding normal distribution; and r is an interpolated, equally spaced array \mc{going from the smallest possible bubble radius, which is the size of an individual pixel up} to the size of the box in units of $h^{-1} \rm{Mpc}$, given by:

\begin{equation}
    \mathrm {BSD \mathrm (A, \mu,\sigma, r)} = \mathrm{\dfrac{A}{r\sigma\sqrt(2\pi)} \exp\left(-\dfrac{1}{2}\dfrac{(\log(r)-\mu)}{\sigma}^2\right)}
\label{eq:BSDparam}
\end{equation}

It should be noted that this parametrization is the best possible representation of the BSD though it is not an absolutely perfect fit. As there can be different size distributions corresponding to same volume averaged ionization fractions, it is not straightforward to uniquely map a power spectrum with a log-normal parametrized BSD and $\AVXHII$.  

\begin{figure*}
    \centering
    \includegraphics[width=6in, height=6in]{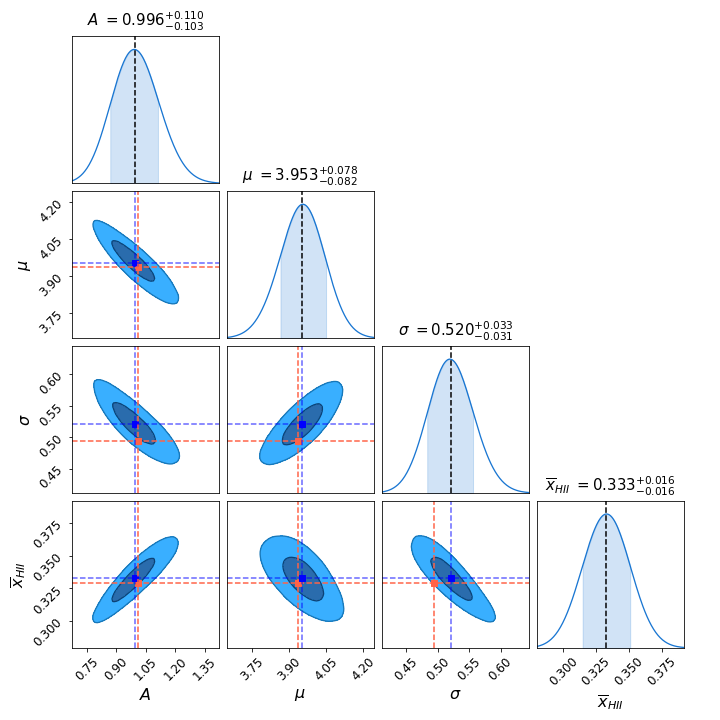}
    \caption{\mcbb{(Updated the figure)}The posterior distributions of the IGM parameters, $A,\mu, \sigma$ (corresponding to the BSD) and $\AVXHII$. The off-diagonal terms show the pair-wise covariances and the diagonal plots show the marginalised distributions. The 2D contour levels represent the $1\sigma$ and $2\sigma$ levels of probability. The shaded regions in the marginalized distributions depict the $68\%$ probability range. The black and orange dashed lines represent the median values of the posterior distribution and the true values of the parameters respectively. }
    \label{fig:mcmc-IGMparameters}
\end{figure*}

We implement a hybrid approach to obtain constraints on the IGM parameters using a Bayesian method as well as an ANN. First we set up our ANN emulator to predict the 21-cm power spectrum, given the bubble size distribution and the ionized fraction (as described in the \textsection\ref{subsection: preparing training sets}). Following this, we employ our power spectrum emulator to a Bayesian sampler, the emulator allows us to avoid a direct dependence on the semi-numerical simulation, thus making this process computationally fast. The final output of the sampler is the posterior probability distribution for the parameters. We report the mode or the most probable value for each of the parameters from the marginalized distributions as the predicted value and calculate the uncertainty bound based on the quantiles. In this work, we implement the ANN emulator into the standard Markov Chain Monte Carlo (MCMC) based parameter estimation method for our data. We use the publicly available code EMCEE to obtain constraints on the 4 IGM parameters, $\rm A, \mu, \sigma, \AVXHII$. Fig.\ref{fig:mcmc-IGMparameters} shows the posteriors obtained and the pair-wise covariances of the IGM parameters for an input 21cm power spectrum which corresponds to an ionization fraction of $0.32$. We obtain good constraints with the combination of Bayesian methods along with the trained ANN-emulator.

\subsection{ANN for parameter estimation}
\label{results:param est}
While we obtain reasonable constraints on the IGM parameters from our Bayesian analysis, we bring to note that not all BSD's can be perfectly parametrized by a log-lognormal distribution. Hence the scope of introducing an error in our estimation already stems from the choice of the parametrization for the size distributions. 

In this subsection, we present an ANN-IGM parameter estimator(ANN-IGMParam), which is trained on the same training set as described in \textsection~\ref{subsection: preparing training sets}. This ANN is also a multilayer perceptron network, with two hidden layers, each with an activation function `elu'. The input layer has 8 neurons corresponding to the input 21-cm power spectrum. The output layer has four parameters, three corresponding to the BSD parameters $(A,\mu, \sigma )$ and the volume-averaged ionization fraction $\AVXHII$. Given the 21-cm power spectrum, we can use this ANN to predict the IGM parameters that describe the EoR.
\begin{figure*}
    \centering
    \includegraphics[width=\linewidth]{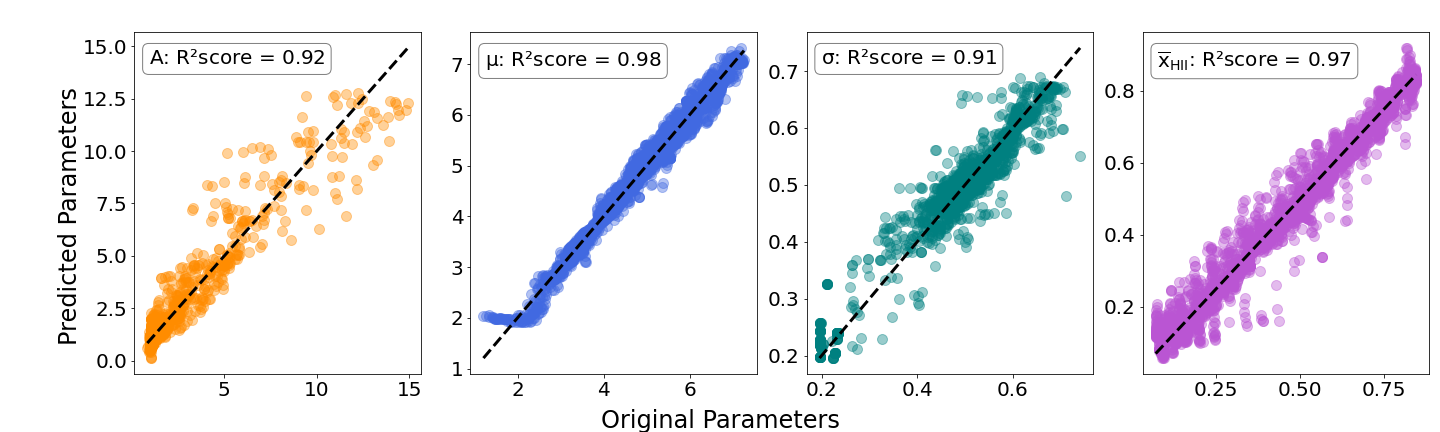}
    \caption{Results from the ANN-IGM parameter estimator(ANN-IGMParam), which is trained to predict the IGM parameters (the BSD parameters $A,\mu,\sigma$ and the ionization fractions $\AVXHII$), given the 21-cm power spectrum. The plots of the true vs predicted values for each of the IGM parameters are shown. The black line in each plot represents the true values. The respective R2 scores are mentioned in the top-left-hand corner of each plot.  }
    \label{fig:ann-IGMparameters}
\end{figure*}

\begin{figure}
    \centering
    \includegraphics[scale=0.4]{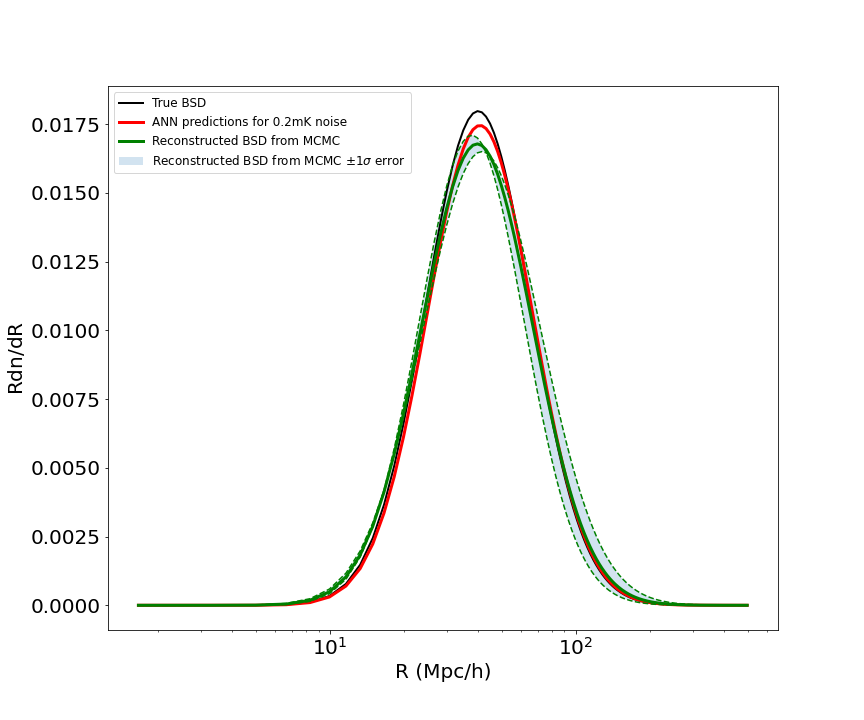}
    \caption{\mcbb{(Updated the figure)}The reconstructed bubble size distributions from the predicted BSD parameters obtained from Bayesian method (green) along with the $1\sigma$ error limits and ANN method (red). The black curve is the BSD with the original parameters.}
    \label{fig:ANN-bayesian}
\end{figure}
\begin{table}
\centering
\begin{tabular}{|c|c|c|c|}
     \hline
     Parameter & True values & Bayesian method & ANN method \\
     \hline
     $\mathrm{A}$ & 1.008 & \mcbb{$\mathrm{0.996}^{+ 0.11}_{-0.10}$} & 0.999\\
     $\mathrm{\mu}$ & 3.935 & \mcbb{$\mathrm{3.953}^{+ 0.078}_{-0.082}$} & 3.954\\
     $\mathrm{\sigma}$& 0.493 & \mcbb{$0.52  ^{ +0.033}_{-0.031}$} & 0.495\\
     $\AVXHII$ & $0.329 $ & \mcbb{$0.33 ^{+ 0.016}_{-0.016}$ }& 0.312 \\
     \hline
\end{tabular}
\caption{A comparison of values of the IGM parameters obtained from the Bayesian and ANN methods, along with the true values. This is one example for comparing the predictions using ANN and Bayesian methods. We find that the ANN based estimator 
outperforms the Bayesian methods, even though the ANN predictions do not come with error contour levels. The corresponding reconstructed BSDs are shown in the Fig.\ref{fig:ANN-bayesian}. }
\label{Tab:2}
\end{table}
To demonstrate the performance of the ANN-IGMParam estimator, we have two test sets:

\begin{enumerate}
\item \textit{Predictions of the IGM parameters from noise-free {\sc grizzly} test sets}\\
This test set consists of a set of 21-cm power spectra from the {\sc grizzly} simulations. Given any 21-cm power spectrum, the ANN predicts the corresponding BSD parameters $(A, \mu, \sigma)$ and $\AVXHII$. We have shown the original versus the predicted values of the parameters from the {\sc grizzly} test set, in Fig.\ref{fig:ann-IGMparameters}.  The R2-scores of the parameters for $A, \mu, \sigma, \AVXHII$ are 0.92, 0.98, 0.91, 0.97 respectively. We observe that the parameter A, which corresponds to the amplitude of the BSD not as well constrained as the other parameters. 

In Tab.\ref{Tab:2} we compare the predictions obtained from the ANN with the results obtained from the Bayesian analysis.  While the ANN takes $0.012$seconds to predict the parameters, the standard Bayesian MCMC method takes $64.457$minutes to converge on a personal computer with 8 cores and 16GB memory.  We find that the results are comparable to the true values of the parameters. Once we have the predicted values of the BSD parameters from the ANN, we use them to reconstruct the BSD using Eq.\ref{eq:BSDparam}. In Fig.\ref{fig:ANN-bayesian}, we show the reconstructed bubble size distributions using the predicted parameters from both methods, along with shaded region showing the $1\sigma$ error on the Bayesian predictions. 

\item \textit{Predictions of the IGM parameters from test sets with SKA noise}\\
We consider a more realistic test set, which contains a set of 21-cm signals contaminated with SKA thermal noise. We use the publicly available code {\sc 21cmsense}\footnote{\url{https://github.com/rasg-affiliates/21cmSense}}(\cite{Pober_2013, Pober_2014}), to generate the sensitivities corresponding to SKA1-low. The sensitivity of an instrument indicates the ability of the instrument to detect the faintest signal possible. In the method followed in {\sc 21cmsense}, a $\textit{uv}$ coverage of the observation is generated by gridding each baseline into the $\textit{uv}$ plane and including the effects of the Earth's rotation over the course of the observation.  

The upcoming SKA telescope would possess sufficient signal-to-noise required to detect the statistical 21-cm signal (\cite{Koopmans_2015}) and also perform 21-cm tomography (\cite{Mellema_2015}).
With 38 m diameter corresponding to a collecting area of $\sim 600~\mathrm m^2$, SKA1-low will have a field of view of about 12.5 $\mathrm {deg^2}$ at 150 MHz. The full SKA1-low configuration will extend up to a diameter of approximately 40 km spread around the core and the longest baseline would be about 65 km. In our calculations, we have used 224 core elements spread within a radius of 500 m. The shortest and longest baseline of the core is 35.1 m and 887 m, respectively. This results in an angular resolution of $10^{'}$ at 150 MHz. We have obtained the thermal noise power spectra for these experiments corresponding to a total of 1080 hours of observation for SKA (in 6 hours of tracking mode observation per day, for 180 days).
Using these thermal noise power spectra, we generate the SKA-noise test set samples. 
\end{enumerate}
\begin{figure*}

\includegraphics[scale=0.30]{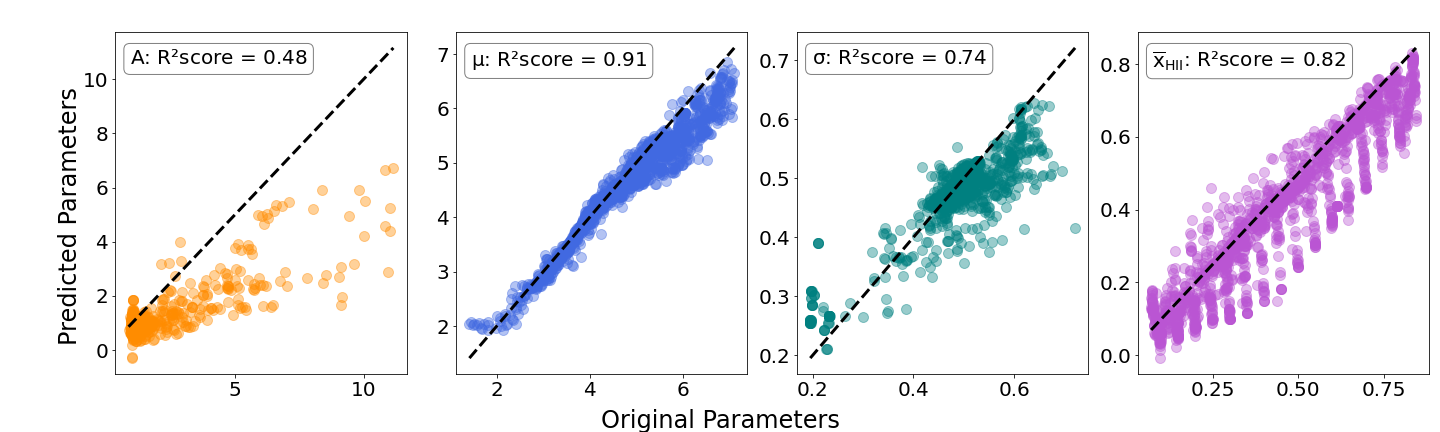}
    \caption{In this figure, we show the scatter plots of true versus predicted IGM parameters $A, \mu,\sigma, \AVXHII$ from the SKA-noise test sets. In the top left-hand corner of each plot, we mention their R2-scores respectively.}
    \label{fig:ska-pred}
\end{figure*}

\begin{figure*}
    \centering
    \includegraphics[scale=0.5]{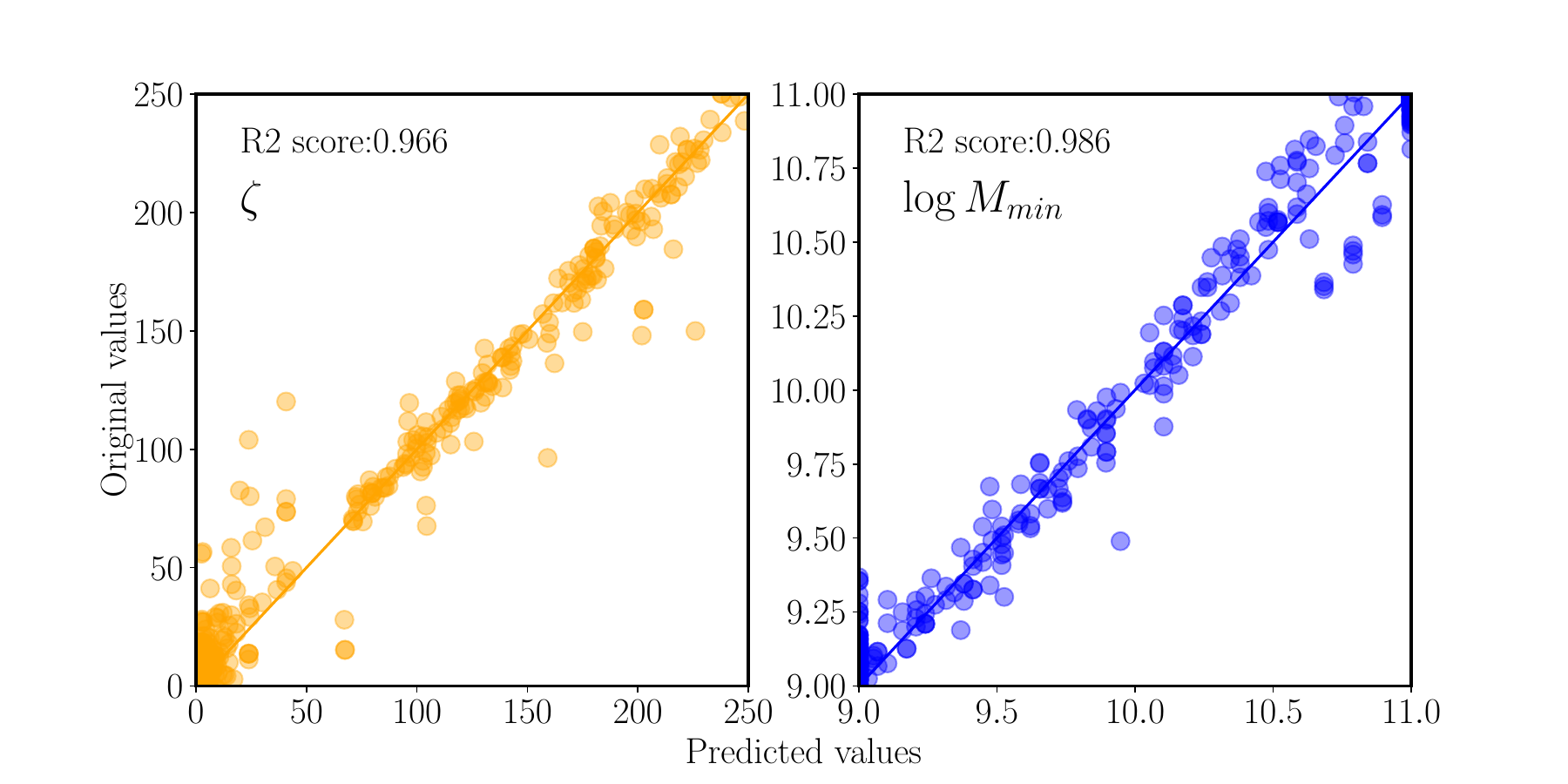}
    \caption{Results of the ANN trained to predict the source parameters from the bubble size distribution and the ionization fractions. The plots of the true vs predicted values for each of the parameters are shown. }
    \label{fig:ANN-source_parameters}
\end{figure*}

 \begin{figure*}
    \centering
    \captionsetup[subfloat]{farskip=2pt,captionskip=1pt}
    \subfloat {\includegraphics[scale=0.3]{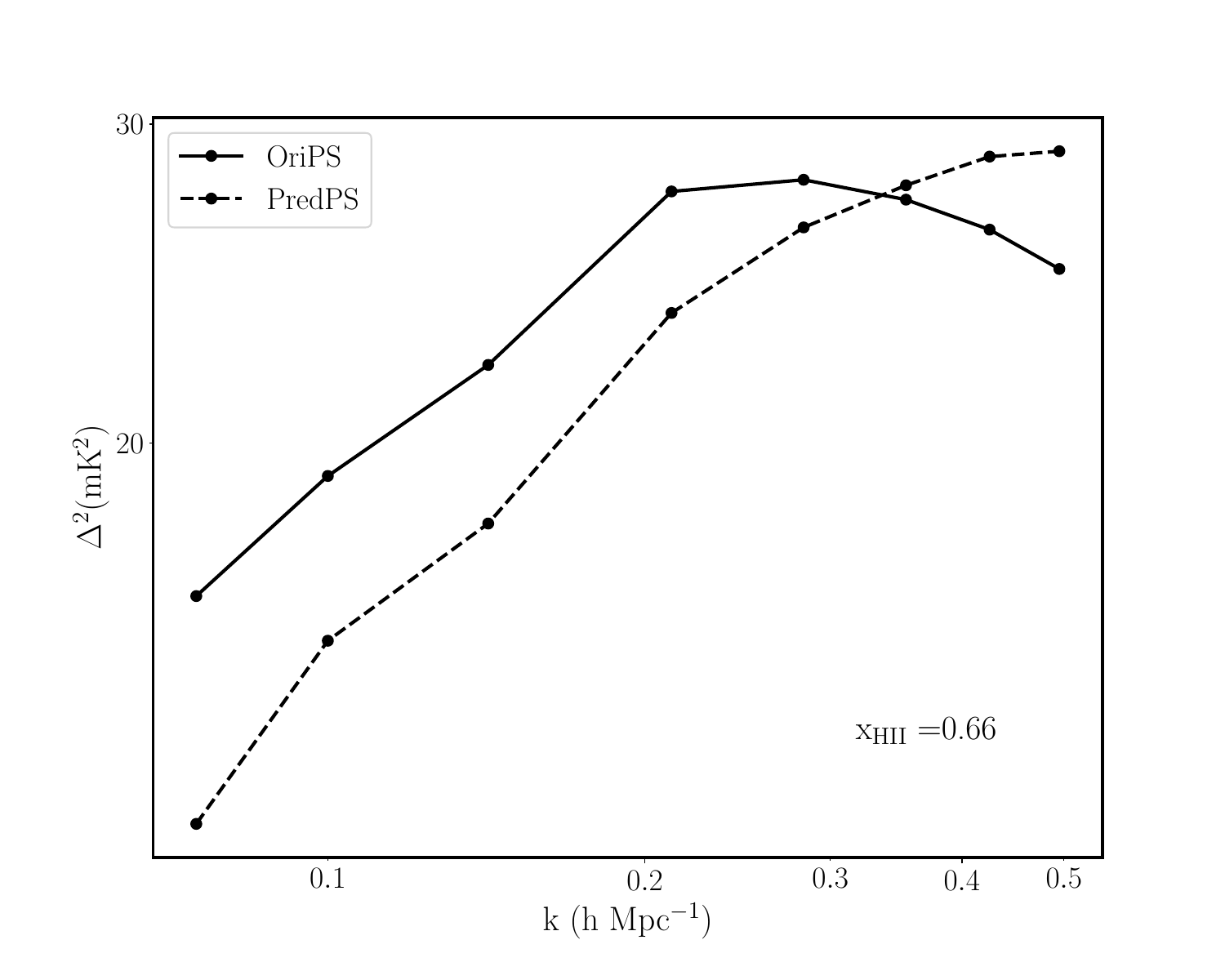}}\\[-5ex]
    \subfloat{\includegraphics[scale=0.3]{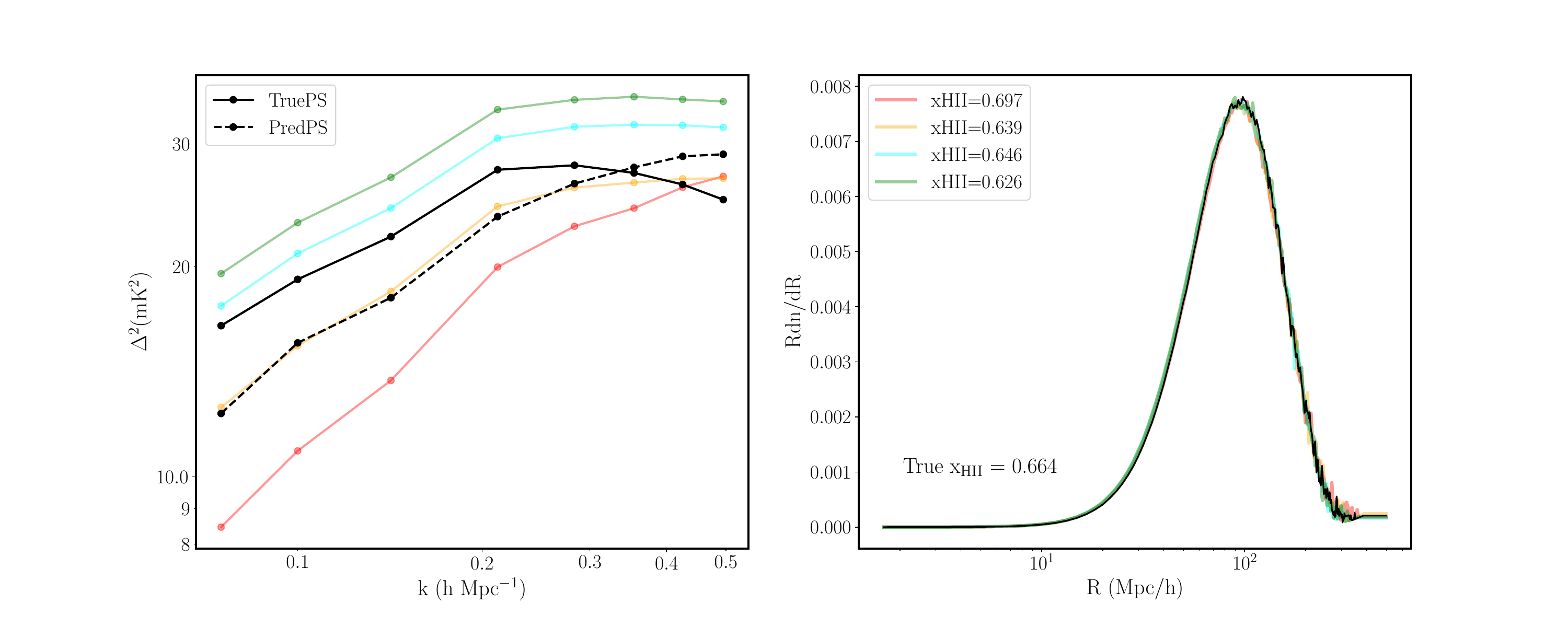}}\\[-5ex]
    \subfloat{\includegraphics[scale=0.3]{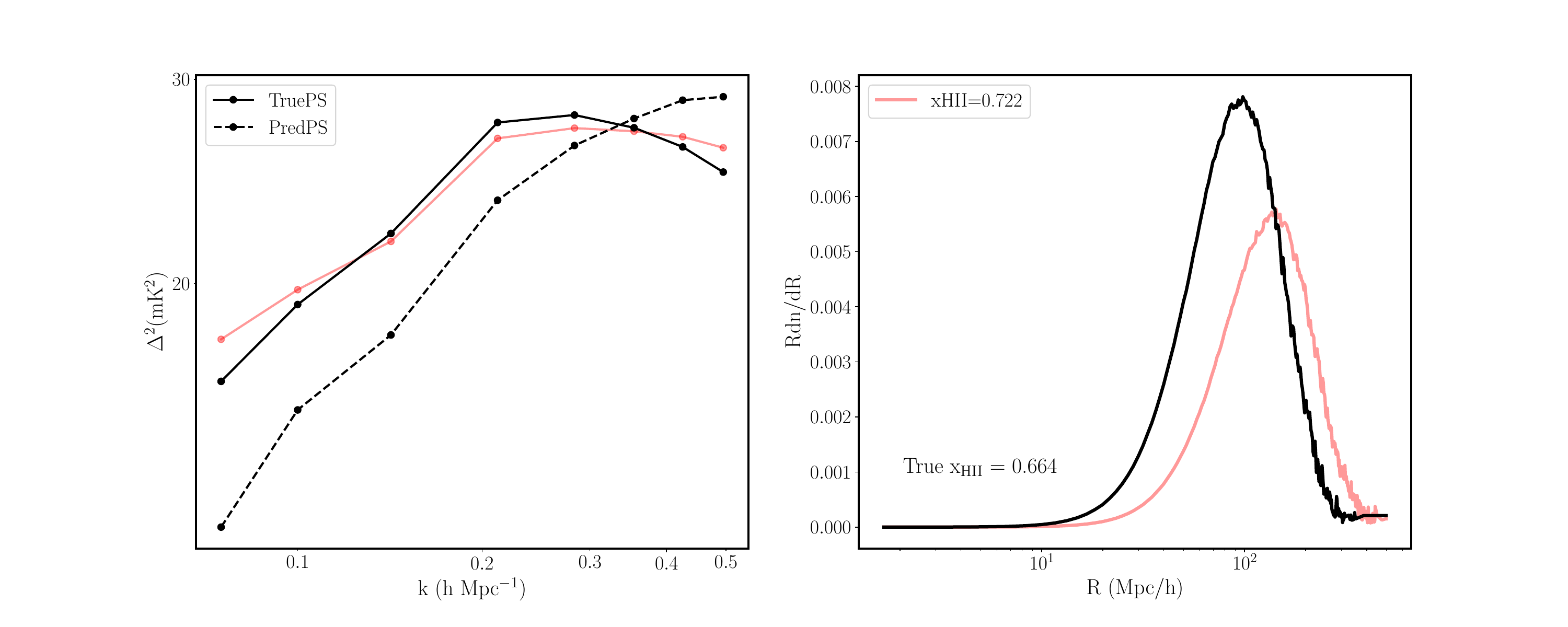}}
    \caption{A crude representation of how crucial a unique mapping of the BSDs and $\AVXHII$ parameters are to the 21-cm power spectrum in the training set. Here we show one test set sample from {\sc  c2ray} simulations. In the first plot, we show the true versus predicted 21-cm power spectrum of the sample from the {\sc c2ray} test set. In the second panel, we see that even though the BSD is quite close to the samples we have trained the ANN with, the corresponding $\AVXHII$ and the 21-cm power spectra are very different. This does not allow the ANN to predict the 21-cm power spectrum as accurately as expected. In the third panel, we search for a power spectrum that is closest to the test power spectrum and plot it. We observe that the corresponding BSD and $\AVXHII$ is very different from the BSD of the {\sc c2ray} sample.}
    \label{fig:c2ray}
\end{figure*}

\begin{figure*}
    \centering
    \captionsetup[subfloat]{farskip=2pt,captionskip=1pt}
    \subfloat {\includegraphics[scale=0.3]{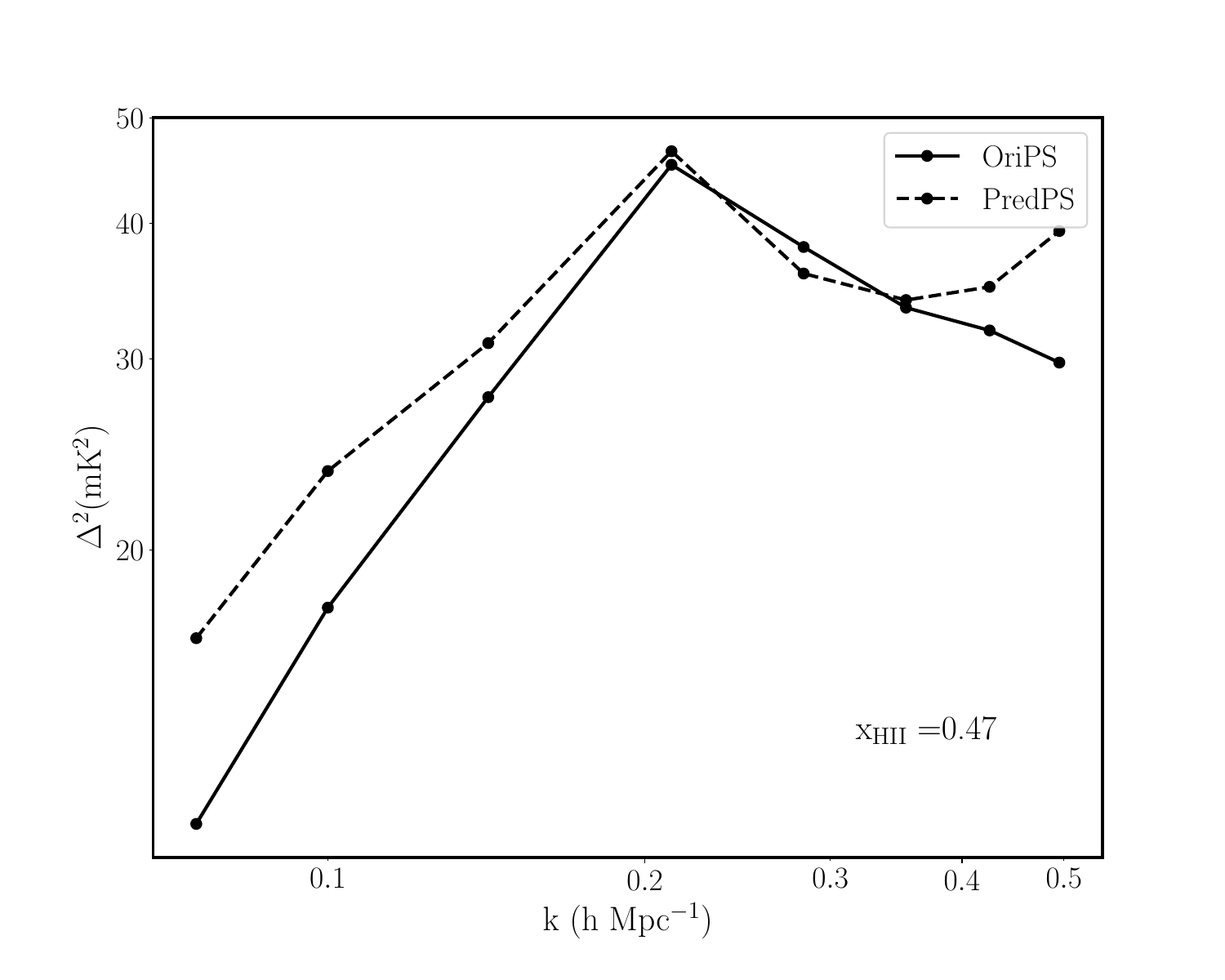}}\\[-5ex]
    \subfloat{\includegraphics[scale=0.3]{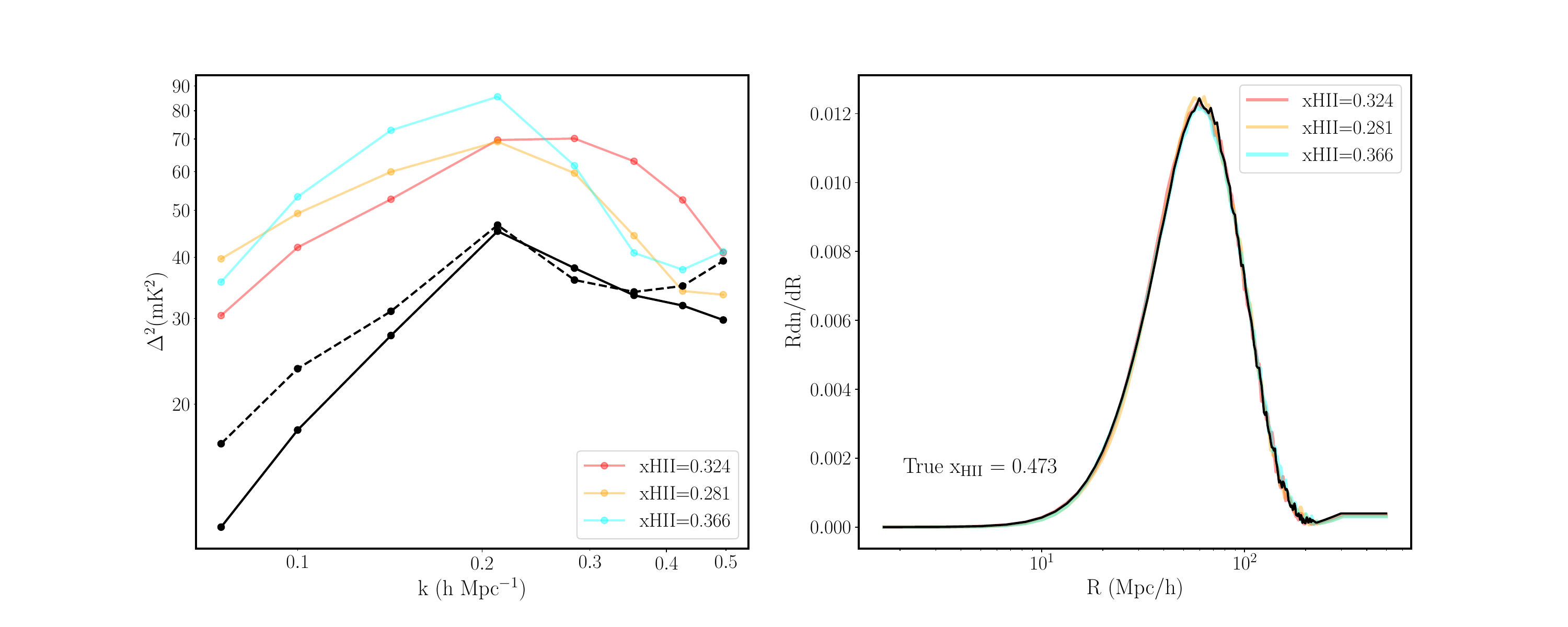}}\\[-5ex]
    \subfloat{\includegraphics[scale=0.3]{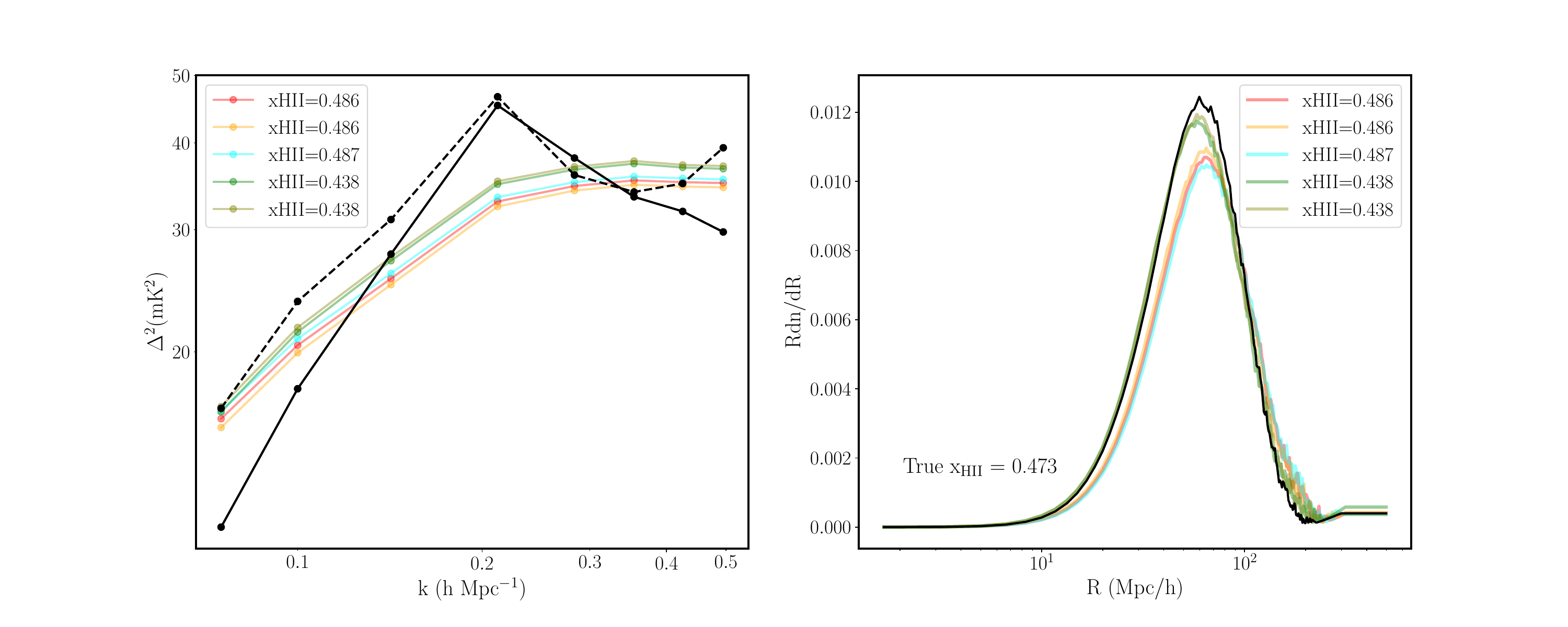}}
    \caption{A crude representation of how crucial a unique mapping of the BSDs and $\AVXHII$ parameters are to the 21-cm power spectrum in the training set. Here we show an example from the {\sc 21cmfast} simulations. In the first plot, we show the true versus predicted 21-cm power spectrum of a sample from the {\sc 21cmfast} test set. In the second horizontal panel, we see that even though the BSDs are close to the samples we have trained the ANN with, the corresponding ionization fractions, $\XHII$ are different. This does not allow the ANN to predict the 21-cm power spectrum as accurately as expected. In the third panel, we search for a power spectrum that is closest to the test power spectrum and plot it. We observe that the corresponding BSD and $\AVXHII$ are very different from what we provide as an input. }
    \label{fig:21cmfast}
\end{figure*}
In Fig.\ref{fig:ska-pred}, we show the original values versus predicted values of each of the IGM parameters from the SKA test sets. We see that the R2-scores are lower than the test set without any SKA-noise. The R2-scores for $A, \mu, \sigma, \AVXHII$ for SKA test sets are respectively 0.48, 0.91, 0.74, 0.82. We observe that when input 21-cm power spectrum is contaminated with SKA noise, the ANN is not able to accurately predict the BSD parameters accurately. However,in comparision there is a smaller effect on the prediction of the volume-averaged ionization fraction from the SKA test set. It should be noted that the ANN-IGMParam has not been trained on the noisy data.

\subsection{Determining the source properties from the IGM properties}
\label{results: source from IGM}
The 21-cm power spectrum is a measurement of the state of the IGM. This in turn does depend on the source parameters but this dependency is indirect and different source parameters may produce similar IGM states. Nevertheless, it is not self-evident how to best parametrize the state of the IGM.
Several recent works (e.g.,\cite{Greig_lofar}) have focused on constraining the source parameters from power spectrum measurements. We would like to stress on the fact that the 21-cm power spectrum has a direct connection with the state of the IGM. Evidently, there is an interplay of several different factors which is neither captured directly by the power spectrum nor can be completely described by the current state of parameters characterizing the IGM. In this paper, we use the size distribution of the ionized regions and the volume-averaged ionization fraction of the IGM as the primary parameters to describe the IGM, assuming no spin temperature fluctuations at redshift 9.1. If we extend our formalism to other redshifts, and include spin temperature fluctuations, heating and feedback effects, then the parameter set describing the IGM would be much larger and complicated. 

As we demonstrated the IGM parameter estimation in the previous sections, we now would like to understand if it is possible to retrieve the source parameters from these IGM parameters directly in the simplest scenario. This framework will be useful if we are able to infer the IGM parameters from an observed 21-cm power spectrum, and would want to understand the nature of the associated source parameters directly from them. We could then compare the inferences about the source parameters from other methods which directly use the 21-cm power spectrum and understand the nature of the sources in a more comprehensive way.\\

For this task, we train another ANN (we refer to as ANN-Source) which takes the IGM parameters as inputs and predicts the source parameters. We use a multilayer perceptron model of ANN which has $301$ neurons in the input layer corresponding to the IGM parameters $(\rm BSD, \AVXHII)$, followed by three hidden layers with activation functions of `elu'.The output layer has two neurons, corresponding to the two source parameters, minimum mass of halos $\mathrm {\log M_{min}}$ and ionizing efficiencies, $\zeta$, which are most relevant for reionization.
We use a training set comprising of 7680 samples and train an ANN to predict  $(\mathrm {\log M_{min}}, \zeta)$, given the BSD and the $\AVXHII$  at redshift $z=9.1$. 
Fig.\ref{fig:ANN-source_parameters} shows the scatter plot of the true versus predicted values of the source parameters that we obtain from the trained ANN-model.  The R2-scores corresponding to the parameters $\mathrm {\log M_{min}}$ and $\mathrm \zeta$, are $0.966$ and $0.986$ respectively, which imply that the predictions are quite close to the original/true values. This is for the first time that we demonstrate such a non-analytical association between the IGM quantities and the source parameters using ANN. Our results suggest this may be interesting to explore further for example for different source models.

\section{Discussions}  
\label{discussions}
The parameter space for the EoR is quite large, and we still do not know with certainty what would be the best set of parameters which would most closely describe the state of the IGM. This is a very difficult task, especially because of the dynamic nature of the evolution of the IGM, which is in turn affected by several unconstrained astrophysical and reionization parameters. Hence, simulation based parameter inference methods (analytical, numerical or ML based) become dependent on the choice of the parameter space, source models and the models of reionization considered. 

We have considered the BSD and  $\AVXHII$ as the parameters to describe the IGM. One of the primary concerns of using a BSD to describe the state of the IGM is the fact that it does not work well in scenarios where there are overlapping bubbles or irregularly shaped ionized bubbles. As the method of defining the bubble size distribution using the mean-free path method is somewhat incomplete (as it does not take into consideration the information of the locations and the shape of the bubbles), it becomes difficult to completely and accurately represent the distribution of the ionized structures. \mcb{In \cite{Lin_2016}, the authors have described various existing methods of characterising the distribution of ionized bubbles, each with their advantages and disadvantages, eventually pointing to the lack of completely understanding of the characteristic scales of the reionization process. } 

The most crucial aspect of implementing supervised learning methods in ML is the choice of the training set. The training set needs to be chosen in such a way that it sufficiently represents all possible reionization scenarios which we want our network to learn about. While this is quite straightforward and simulation model dependent, when one considers designing an emulator to map 'controllable' EoR source parameters to the 21-cm power spectrum, it becomes slightly complicated while we try to map the derived IGM parameters to the 21-cm power spectrum. As different kinds of morphologies can result in the same volume averaged ionization fraction, there is no unique mapping between a combination of a given size distribution and the volume averaged ionization fraction, and the associated 21-cm power spectrum. This also makes it complicated to have an ideally uniformly sampled input parameter space. In this work, we included several possible reionization scenarios leading to various shapes of power spectra. For the ANN emulator, the input parameter space is comprised of the corresponding BSD and $\AVXHII$ and the target space consists of the the 21-cm power spectra. This work demonstrates the training of an ANN with a combination of a continuous distribution(BSD) and discrete ($\AVXHII$) values in the input parameter space.

Another common issue with ML methods when used with simulations, is the problem of generalizability. We observe that in supervised learning methods, the ANN performs poorly, if the test sets are derived from an entirely different simulation. This could be due to the underlying differences in the implementation of the physics models and various other details and approximations.
 
To test how well the emulator could predict the 21-cm power spectrum from a test set which is not generated by the {\sc grizzly} simulation, we use the simulations from the 3D RT code, {\sc c2ray}. In this example, we take a particular sample from the {\sc c2ray} test set and input its IGM parameters, BSD and $\AVXHII$ to obtain the predicted 21-cm power spectrum. In the top centre plot in Fig.\ref{fig:c2ray}, we show true versus predicted 21-cm power spectra using the {\sc c2ray} test set. In the following two horizontal panels, we demonstrate how the absence of similar models in the training set affects the prediction from the ANN for the power spectrum plotted in pink. In the second row plots, we look for the closest match between the test set {\sc c2ray} sample and the samples in the training set. \mcb{To identify the best-matching test sample in terms of the bubble size distribution (BSD) and power spectrum from the training set, we compute a metric based on minimizing a mean-squared error, $E$. For the BSDs, $E_{\rm BSD}^i$ is defined as: $    E_{\rm BSD}^i= \frac{1}{300} \sum_{R}\left(\rm BSD_{R}-\rm BSD_{test,R}\right)^2$, where the summation runs over all 300 $R_{\rm bins}$ considered in the calculation of the BSDs. We compute $E_{\rm BSD}$ for each sample, $i$, in the training set and identify the sample with the smallest $E_{\rm BSD}$ as the closest match to the test sample. This serves as an assessment of whether a test sample from {\sc c2ray} or {\sc 21cmFast} has a close counterpart in our training set. The corresponding power spectra and their ionization fractions are plotted in the second panels of Fig.\ref{fig:c2ray} and Fig.\ref{fig:21cmfast}.}

We find that there are a few BSDs which match the {\sc c2ray} BSD with a \mcb{ $E_{\rm BSD}\sim 10^{-7}$.} Plotting only the BSDs which have ionization fractions in between $0.60-0.70$, we observe that these BSDs correspond to quite a large range of shape and amplitudes in the 21-cm power spectrum space, particularly towards the lower $\mathit{k}$-modes. 

 \mcb{Similarly, we look for the closest match of the power spectra in the Training set I, and we do so by computing a $E_{\rm \Delta^2}^i$ for the power spectra as: 
$E^i_{\Delta^2}= \frac{1}{8} \sum_{k}\left(\Delta^2_{k}-\Delta^2_{{\rm test}, k}\right)^2$,
where, the summation is over the 8 k-bins and a $E_{\rm \Delta^2}$ is calculated for each sample, $i$, in the training set.} We find that the closest match available has a corresponding \mcb{$E_{\rm \Delta^2}\sim 0.02$}, but the BSD and $\AVXHII$ of that particular case is very different from the {\sc c2ray} sample being considered.
  
We show another similar test case, using a test set sample from the seminumerical simulation, {\sc 21cmfast} in Fig.\ref{fig:21cmfast}. We find that there is no sample in our training set which represents the kind of power spectrum that is expected from the {\sc 21cmfast} test set sample. 
\mcb{This check confirms that our training set does not contain close matches to the {\sc 21cmFast} and {\sc c2ray} test samples. As a result, the neural network struggles to generalize to these simulations, highlighting the need for a more diverse and representative training set to capture a broader range of reionization morphologies.} This also implies that each of the simulations have very different approximation techniques and non-generalizable models which plays a crucial role in determining the distribution of ionized sources and the power spectrum.

To make the training set most diverse and well-represented, we incorporated different shapes of the 21-cm power spectrum corresponding to the {\sc c2ray} models in our training sets.  We observed that even if we are able to match the shape of the power spectrum by making the ionization model more extreme by using a different power-law model (for example, using higher values of $\alpha$, as discussed in \textsection~\ref{subsection: preparing training sets}), the corresponding BSDs are not similar to those obtained from {\sc grizzly}. We demonstrate the crucial association of the IGM parameters, the BSD and $\AVXHII$ with the shape of the power spectrum in Fig.\ref{fig:c2ray} and Fig.\ref{fig:21cmfast}. We realize that our training set is not sufficient to work seamlessly for test set samples from other simulations which use drastically different models as compared to {\sc grizzly}. We plan to incorporate more diverse models 
and develop a generalizable and complete training set in our future work.

Another important point which should be noted is the completeness of the IGM parameter set that has been considered in this work. While the BSD and $\AVXHII$ does a fairly good job of describing the power spectrum, we are limited to one kind of simulation only ({\sc grizzly}, with a specific kind of source model as discussed in detail in earlier sections). Our ANN emulator is not accurate when we try to test it on other kind of simulations which have their respective source models, parameters and assumptions. It is important to understand that the {\sc grizzly}-trained network is not successful in inferring the IGM parameters from the 21-cm signal from different simulations, possibly due to different algorithms used in these codes. This could imply that any interpretation of future 21-cm results could be code-dependent, which is a matter of concern. It is important to consider cross-collaborating between different code groups and arrive at a standardized parametrization which would enable us to switch between different codes and incorporate them in our inference frameworks. We are exploring novel methods to incorporate a wide variety of size distributions and different shapes of the 21-cm power spectrum, for future work.

\section{Summary \& Conclusions}
\label{conclusions}

\mc{In this paper, we present a ANN based machine learning framework to extract the IGM parameters (BSD and the $\AVXHII$) simultaneously from the redshifted EoR 21-cm power spectrum for the first time.} We have simulated 21-cm power spectra at redshift $z=9.1$, using {\sc grizzly}, which is a 1D RT code. We have covered two different aspects. Firstly, we have designed an ANN emulator using simulations from {\sc grizzly}, which predicts the 21-cm power spectrum at redshift 9.1, given the size distribution of the ionized bubbles and the ionization fraction. Using this emulator, we have constrained the IGM parameters using a Bayesian framework incorporating the standard MCMC sampling. 

Next, to assess the performance of our emulator based Bayesian framework, we have trained a different ANN for the IGM parameter estimation. We find that the results obtained from using MCMC and the ANN are quite comparable. Both these techniques for parameter estimation come with their pros and cons. The Bayesian method has the advantage of providing us with the confidence levels along with most likely values of the inferred parameters. However, this takes a considerable time. For this work, the bayesian algorithm took about 1 hours to converge on a machine with 40 cores. On the other hand, the ANN predictions take less than a second to predict. From the ANN predictions of the parameters in Fig.\ref{fig:ann-IGMparameters}, we see that the parameters  $\sigma$ and $\AVXHII$ have higher R2 scores, while $A$ and $\mu$ are not that well predicted. 

\mc{The question arises that why we would choose to compare the ANN with MCMC for parameter estimation. We do so because MCMC is widely regarded as the standard method in 21-cm cosmology for both parameter estimation and uncertainty quantification. As the field increasingly explores novel machine learning approaches in parameter inference, including neural ratio estimation and simulation-based inference (SBI), our intention was to highlight the differences in computational efficiency and applicability between a traditional approach (MCMC) and a faster, alternative prediction framework (ANN).
We acknowledge that MCMC’s strength lies in sampling complex, high-dimensional posterior distributions rather than simply identifying maximum-likelihood estimates. However, in this case, we aimed to show the feasibility of ANNs in obtaining accurate parameter predictions—though, as noted, this does not yet capture the full posterior uncertainties that MCMC provides.}

However, when we test our Bayesian framework implementing the ANN-emulator for non-{\sc grizzly} test sets, it does not perform well.
With test set from other simulations, like {\sc c2ray} and {\sc 21cmfast}, the accuracy of prediction dropped drastically. The possible explanation of this could be the difference in the implementation of the source models in these simulations, and other underlying limitations and approximations. We found that the mapping between 21-cm EoR power spectra and their corresponding [BSD, $\AVXHII$] were very different from the training set, when we considered test sets from other simulations. 

In the final part of the paper, we demonstrate the use of ANN to predict the source parameters directly from the IGM parameters. We obtain very high accuracy for the predicted source parameters. Through this approach, we are able to demonstrate the feasability of an ANN method of inference of the source parameters directly from the IGM parameters. 

In conclusion, using artificial neural networks in inferring IGM and source parameters is accurate and fast. Even though we are limited by the scope of the training sets in supervised learning methods, we can achieve very high accuracies in a very short timescales. In the future, we can make the training sets more informative so that it works well with any kind of simulations and becomes more generalizable. We plan to also extend this framework to include a range of redshifts spanning CD and EoR. Such a complementary approach using ML will be useful in identifying and breaking degeneracies in model selection and parameter estimation methods with regard to EoR 21-cm power spectrum. 
\mc{Another interesting aspect to explore as a follow-up to this work would be to rigorously account for the uncertainties propagated throughout the  inference framework, at different levels. We plan to systematically investigate the effects of introducing emulator errors, errors introduced by a different parametrization, in addition to errors and systematics associated with the observing instrument. }


\acknowledgments
MC acknowledges the computing resources provided by ARCO, Israel and the CFPU postdoctoral fellowship.  MC, RG, AKS and SZ acknowledge support from the Israel Science Foundation (grant no. 1388/24). AKS also acknowledges support from National Science Foundation (grant no. 2206602). LVEK acknowledges the financial support from the European Research Council (ERC) under the European Union’s Horizon 2020 research and innovation programme (Grant agreement No. 884760, "CoDEX"). GM’s research has been supported by Swedish Research Council grant $\rm {2020-04691\_VR}$. RG acknowledges support from the Kaufman Foundation (Gift no. GF01364) and support from SERB, DST Ramanujan Fellowship no. RJF/2022/000141.


\bibliographystyle{JHEP}
\bibliography{total}






\end{document}